\documentclass{article}
\usepackage{forest}
\usepackage{setspace}
\usepackage{biblatex} 
\usepackage{amsmath}
\RequirePackage{fullpage}
\RequirePackage[most]{tcolorbox}
\RequirePackage{chngcntr}
\RequirePackage{amsmath, amsthm, amsfonts, amssymb}
\RequirePackage[hidelinks]{hyperref}
\RequirePackage{graphicx}
\RequirePackage{comment}
\RequirePackage{subcaption}
\RequirePackage{caption}
\RequirePackage{enumitem}
\RequirePackage{physics}
\RequirePackage{tikzlings}
\RequirePackage{comment}
\RequirePackage{tikz}
\usetikzlibrary{tikzmark}
\usetikzlibrary{calc, karnaugh, arrows, circuits.logic.US, angles, quotes,fit,positioning}
\usetikzlibrary{quantikz}
\usepackage{kvmap}
\usepackage{pstricks,multido}
\psset{dimen = m, fillstyle = solid}
\usepackage{xcolor}
\usepackage[margin=1in]{geometry}
\usepackage{blindtext}
\usepackage{subfiles} 
\usepackage{ragged2e}
\usepackage{changepage}
\usepackage [english]{babel}
\usepackage [autostyle, english = american]{csquotes}
\MakeOuterQuote{"}
\usepackage{float}

\title{\vspace{-1.2cm}Synthesis of Binary-Input Multi-Valued Output Optical Cascades for Reversible and Quantum Technologies}

\begin{document}
\date{\vspace{-9ex}}

\maketitle
    \begin{center}
        Ishani Agarwal
        
        ishani2807@gmail.com
        
        Department of Electrical and Computer Engineering, Portland State University 

        Portland, Oregon 97201, United States

    \end{center}

    \begin{center}
        Miroslav Saraivanov 

        michaelsaraivanov@yahoo.com
        
        Department of Electrical and Computer Engineering, Portland State University 
        
        Portland, Oregon 97201, United States
    \end{center}

    \begin{center}
        Marek Perkowski 

        h8mp@pdx.edu
        
        Department of Electrical and Computer Engineering, Portland State University 
        
        Portland, Oregon 97201, United States
    \end{center}

\vspace{-2ex}

\[\textbf{Abstract}\]
\begin{adjustwidth}{1.7cm}{1.7cm}
\justify{This paper extends the decomposition from the group theory based methods of Sasao and Saraivanov to design binary input multivalued output quantum cascades realized with optical NOT, SWAP, and Fredkin Gates. We present this method for 3, 5, and 7-valued outputs, but in general it can be used for odd prime-valued outputs. The method can be extended to realize hybrid functions with different valued outputs. A class of local transformations is presented that can simplify the final cascade circuits. Using these simplifying transformations, we present an upper bound on the maximum number of gates in an arbitrary $n$-variable input and $k$-valued output function.}
\end{adjustwidth}

\section{Introduction}

\indent

The implementation of reversible logic in the computational industry is considered to be one of the most promising solutions to reduce power consumption per logical operation, and one of the major applications of reversible logic is in quantum computing [\ref{32}-\ref{34}]. Continuing to increase the capacity to transmit and process both quantum and classical information is necessary in the computational industry. A key solution is derived from Landauer’s principle—which was experimentally proved in 2012 [\ref{50}]—which stipulates a minimum energy dissipation for each irreversible operation [\ref{36}, \ref{51}]. Achieving Landauer’s limit necessitates the use of reversible computation, where inputs can be reconstructed from their outputs. As shown in [\ref{47}], Landauer’s assumption can be broken through optical interference, acting as a reversible element. Hence, reversible computing offers the potential to achieve zero dissipation by preventing entropy loss during computation. Moreover, with computational demand for reduced power per bit operation, we have to fundamentally revise computer design principles, transitioning from electron-based to photon-based systems, as it is originally discussed in [\ref{32}] with generative artificial intelligence models and cryptocurrency mining and exchanging [\ref{64}]. 

Optical systems, particularly integrated photonics, offer distinct advantages over traditional electronics. Photons, being massless and traveling at the speed of light, enable faster data transmission with minimal latency, essential for scaling both quantum and classical systems. Additionally, optical components support high parallelism, improving the efficiency of information processing. Photons generate less heat than electrons, reducing power consumption and cooling requirements, making optical systems energy efficient [\ref{59}].

Furthermore, multivalued logic is said to be the best solution in future all-optical signal processing systems as it can increase the data-carrying capacities, large information storage, and high-speed arithmetical operations [\ref{7}]. With recent emerging technologies such as Rigetti Aspen, multivalued computers are becoming practical as seen in [\ref{30}]. It has already been shown that multivalued logic can efficiently be applied to various optical computation solutions by using the polarization states of light along with its presence or absence [\ref{48}, \ref{49}, \ref{54}, \ref{55}]. Our methodology is designed for optical quantum and non-quantum technologies, not non-optical technologies.

Potentially, there exist four types of optical reversible or optical reversible quantum circuits: (1) Functions with binary input and binary output (2) Functions with binary input and multivalued output (3) Functions with multivalued input and binary output (4) Functions with multivalued input and multivalued output. Several authors have addressed the problem of minimizing quantum circuits with binary inputs and binary outputs, such as [\ref{1}, \ref{3}-\ref{6}] to list just a few. There also exist a few methods to synthesize functions with multivalued input and binary or multivalued output [\ref{17}-\ref{19}, \ref{44}, \ref{5}]. These methodologies were for non-optical technologies.

However, circuits with binary input and multivalued output are the least commonly studied but are the focus of this paper. There is currently no research on the circuit realization of binary input and multivalued output quantum circuits, particularly for optical technologies. We demonstrate our group theory-based methodology using quantum circuits with binary inputs and 3, 5, or 7 output values, but our method can be easily extended to other radices of functions. Group theory was previously used to design cascade circuits in [\ref{2}, \ref{9}, \ref{10}, \ref{16}].

The only gates that we require in the proposed method are binary inverters, single binary-controlled multivalued-targets Fredkin gates (one control of binary logic and two targets of multivalued logic), and multivalued SWAP gates. Current optical computing systems have successfully implemented both binary and multivalued reversible logic and are able to construct circuits using combinations of CNOT and Fredkin gates [\ref{32}, \ref{56}, \ref{29}, \ref{37}, \ref{45}, \ref{46}, \ref{52}, \ref{53}]. In particular, Figure 6 from [\ref{32}] shows the basic realization of a Fredkin gate using a Mach-Zehnder interferometer.

In several other synthesis methods for multivalued quantum circuits as well as in binary quantum circuits, the number of SWAP gates grows rapidly as presented with examples in [\ref{20}]. Because in some modern quantum layouts a qubit, \(A\), has four neighbors, then selecting this qubit as a target and giving variables \(x_1, x_2, x_3, x_4\) as its neighbors, we do not need SWAP gates to control multivalued qubit \(A\) with these four neighbors. Controlling other variables with these four neighbors may be more difficult and require SWAP gates, but still, the total number of SWAP gates is smaller for cascades than for other quantum circuit structures mapped to quantum layouts. While we give here our motivation for multivalued cascades, the problem of layout is not further discussed in this paper. The quantum layout problem is discussed in [\ref{20}, \ref{15}, \ref{23}, \ref{24}] and the physical implementation of multivalued quantum gates is further discussed in [\ref{7}, \ref{11}, \ref{12}]. More specifically, multivalued Fredkin gates can be efficiently realized in several quantum technologies, including optical systems, but this paper is devoted only to optical systems [\ref{32}, \ref{56}, \ref{26}, \ref{27}, \ref{28}, \ref{29}].

The fundamental realization of optical reversible circuits is the Fredkin gate [\ref{32}]. The multivalued Fredkin Gate is a key stand-alone gate in optical computing and has a fidelity of \(99.75\%\) [\ref{32}, \ref{37}, \ref{35}, \ref{38}, \ref{39}, \ref{40}, \ref{41}]. It can be used to design 16 Boolean logical operations and multivalued circuits. As all of the outputs of the 16 circuits have logical states of the 16 Boolean operations, there is no garbage in their design [\ref{7}].

As compared to the M-S gate, which is a purely theoretical gate, in this paper we introduce a method to create quantum circuits only using gates that are experimentally realizable on optical hardware. Yet, note that the ternary Fredkin gate and SWAP gate can also be realized using Muthukrishnan-Stroud gates (M-S gates) [\ref{42}, \ref{43}, \ref{60}, \ref{61}]. Decomposing the SWAP gates into a combination of Fredkin gates and inverters and then decomposing these Fredkin gates into M-S gates, our methodology can be extended to produce circuits that only require M-S gates and binary inverters. However, Fredkin gates are much simpler than M-S gates for higher-level valued logic. Our paper is focused on reversible optical logic and quantum optical logic, with a focus on multivalued logic, so we do not expand on the extension of decomposing the Fredkin gates and SWAP gates into M-S gates in the remainder of the paper.

To the best of our knowledge, no optical technology exists that demonstrates the implementation of a multivalued Toffoli-like gate. Hence, methodologies that synthesize quantum circuits using multivalued Toffoli-like gates are not practical in optical quantum technologies. Our approach is unique because it uses Fredkin as a base in contrast to most other researchers that use Toffoli as a base. An additional advantage of our methodology is that Fredkin and SWAP gates are of rather similar complexity for different radices, so our method remains without modification for higher radices. In contrast, for any radix greater than three, Toffoli gates get increasingly more complicated and are different from one radix to another and must be individually designed by the user. Additionally, ternary Fredkin and quaternary Fredkin are quite different in non-optical technologies, while they are very similar in optical technologies [\ref{32}].

Again, this methodology applies to multivalued optical technologies but not to ternary non-optical technologies because it is quite expensive. The ternary realization of the gates used in this method is cheap in optical technologies but more expensive in other technologies. Our method is applicable to both multivalued reversible non-quantum logic and multivalued reversible quantum logic. Although our method relates directly to reversible non-quantum logic, many quantum algorithms—including Grover's algorithm and quantum walk—have oracles realized with reversible quantum logic. These oracles do not use non-permutative gates. However, they are part of complete quantum algorithms that include truly quantum gates such as the Hadamard and phase gates. 

Please observe that in this paper, we employ the notation of quantum gates used in both quantum superconducting and optical domains. There are other notations used specifically for optical domains as in [\ref{62}, \ref{63}].

This paper is organized as follows. In section 2, we introduce group theory and generate groups that will be used in our approach. In section 3, we introduce group function decomposition as well as examples for one, two, and three-variable input circuits. Section 4 lists the local transformations that are used in our methodology. Section 5 presents additional examples of our methodology. Section 6 proves an upper bound on the maximum number of cells and gates in the canonical cascade and quantum circuit respectively in an arbitrary binary input and multivalued output function using our method.

\section{Group Theory}

A binary operation is an operation performed on two elements of a set to obtain a third element in the set. A group, denoted $\langle G, \star \rangle $, is an algebraic structure consisting of a set $G$ with a binary operation, $\star$, that satisfies the four properties of closure, identity, associativity, and invertibility.  

\begin{enumerate}
    \item Closure: $a \star b \in G$ for all $a,b \in G$.
    \item Identity: There exists a unique $e \in G$ such that $e \star a = a \star e = a$ for all $a \in G$.
    \item Associativity: $a \star (b \star c) = (a \star b) \star c$ for all $a,b,c \in G$.
    \item Invertibility: For all $a \in G$, there exists $b \in G$ such that $a \star b = b \star a = e$ where $e$ is the identity element.
\end{enumerate}

The order of a group is equal to the number of unique elements in its set. An Abelian group, $\langle G, \star \rangle $, is a commutative group that satisfies $g_1 \star g_2 = g_2 \star g_1$ for all $g_1, g_2 \in G$. A cyclic group is a group that is generated by one of its elements, $g$. For example, the group $\langle g^0, g^1, g^2, g^3 \rangle $ where $g^4 = g^0$ is a cyclic group of order 4. The dihedral group, denoted $D_n$, is a non-Abelian group of order $2n$ consisting of the rotational and reflectional symmetries of a regular $n$-gon. In general, $D_n =  \langle a^0, a^1 \ldots a^{n-1}, g, ag, a^2g \ldots a^{n-1}g \rangle $ where $a^n = a^0$ and $g^2 = g^0$. $D_n = C_n \times C_2$ where $C_k$ is the cyclic group of order $k$. In order to use group functions to realize logical functions, we must avoid groups in which two different input combinations can produce the same output value. This means that we need to create groups that are non-Abelian.

Consider the groups of order 2 through 14. In order for our method to be applicable, we require a non-Abelian group with two variables and an odd valued output as shown in Figure \ref{f1}. We present a proof for the requirement of an odd valued output in section 3.4. 

\begin{figure}[!ht]
    \centering
    \begin{tabular}{c|c|c|c}
   \textbf{Order} & \textbf{Non-Abelian group?} & \textbf{Odd Valued Output?} & \textbf{Our Method Applicable?} \\

    \hline

    2 & No & — & No  \\
    3 & No & — & No  \\
    4 & No & — & No  \\
    5 & No & — & No  \\
    6 & Yes & Yes & Yes  \\
    7 & No & — & No  \\
    8 & Yes & No & No  \\
    9 & No & — & No  \\
    10 & Yes & Yes & Yes  \\
    11 & No & — & No  \\
    12 & Yes & No & No  \\
    13 & No & — & No  \\
    14 & Yes & Yes & Yes 
    \end{tabular}
    \caption{Applicability of groups of order two through fourteen}
    \label{f1}
\end{figure}

\noindent A proof for the existence of non-Abelian groups in Figure \ref{f1} above can be found in [9, 14]. Hence, based on the table, the possible groups that can be used in our method are the groups of order 6, 10, and 14, or $D_3, D_5, \text{ and } D_7$ respectively, and we will use these groups to demonstrate our methodology.

\subsection{Dihedral Group Generation}
{\parindent0pt

\subsubsection{Group $D_3$}

The group $D_3 = \langle a,g: a^3=g^2 = e, gag = a^{-1} \rangle $. 

The group maps for elements $a, g$ are shown in Figure \ref{f2}. The group maps illustrate the behavior of the elements after composing them with either $a$ or $g$. For example, when $g$ is composed with $g$, the resulting element is $I$.

\begin{figure}[!ht]
    \centering
    \includegraphics[width=8cm]{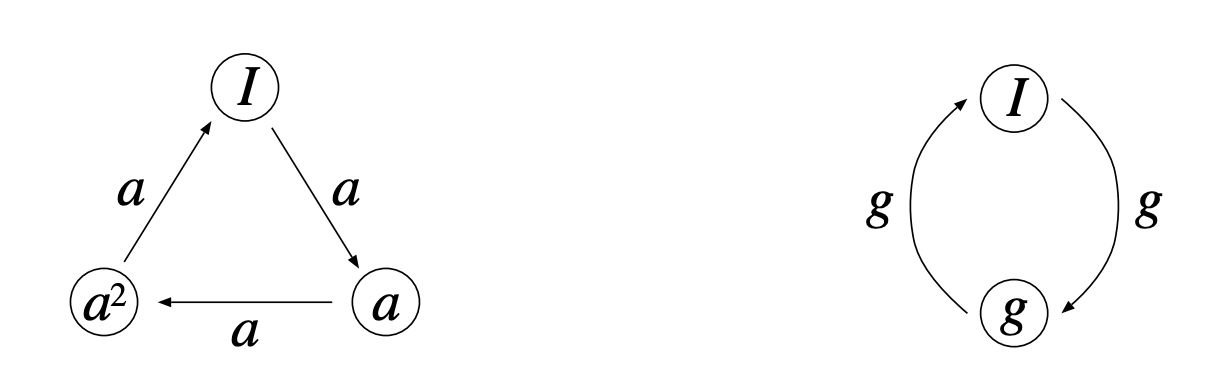}
    \caption{Group Maps for elements $a,g$}
    \label{f2}
\end{figure}
We construct the group $D_3$ using elements $a$ and $g$ with $I$ as the identity element as shown in Figure \ref{f3}. Elements $I, g$ and $a$ each have three horizontal rails where each rail represents a qubit. Element $g$ swaps the bottom two rails, and element $a$ maps the $n$th rail with the $n+1$th rail where the bottom rail maps to the first rail.

\begin{figure}[!ht]
    \centering
    \includegraphics[width=8cm]{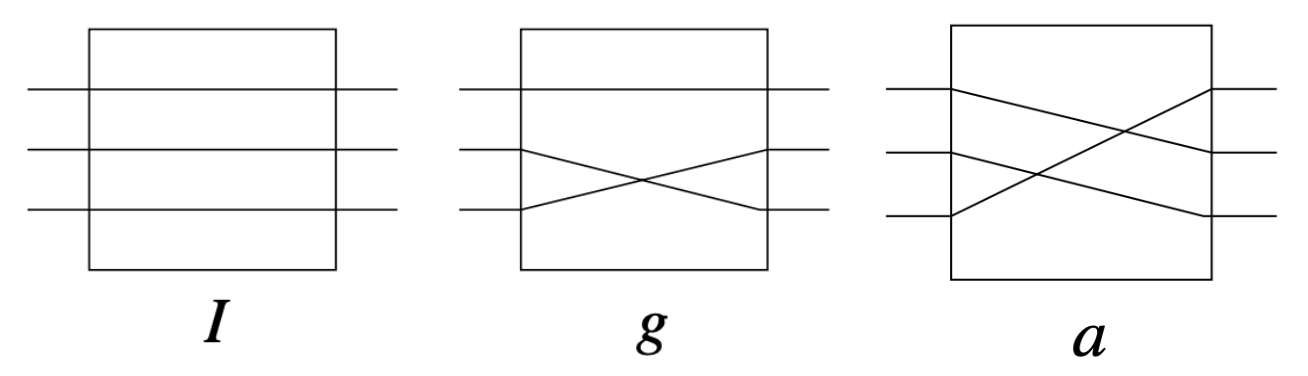}
    \caption{$D_3$ elements $I, g, a$}
    \label{f3}
\end{figure}

We can generate the other elements of this group by composing $a$ and $g$. For example, $a^2$ can be generated as shown in Figure \ref{f4}:

\begin{figure}[!ht]
    \centering
    \includegraphics[width=8cm]{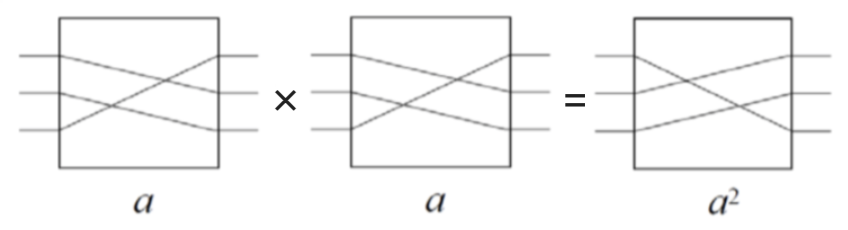}
    \caption{Element $a^2$}
    \label{f4}
\end{figure}

All the different elements of this group are shown in Figure \ref{f5}.

\begin{figure}[!ht]
    \centering
    \includegraphics[width=8cm]{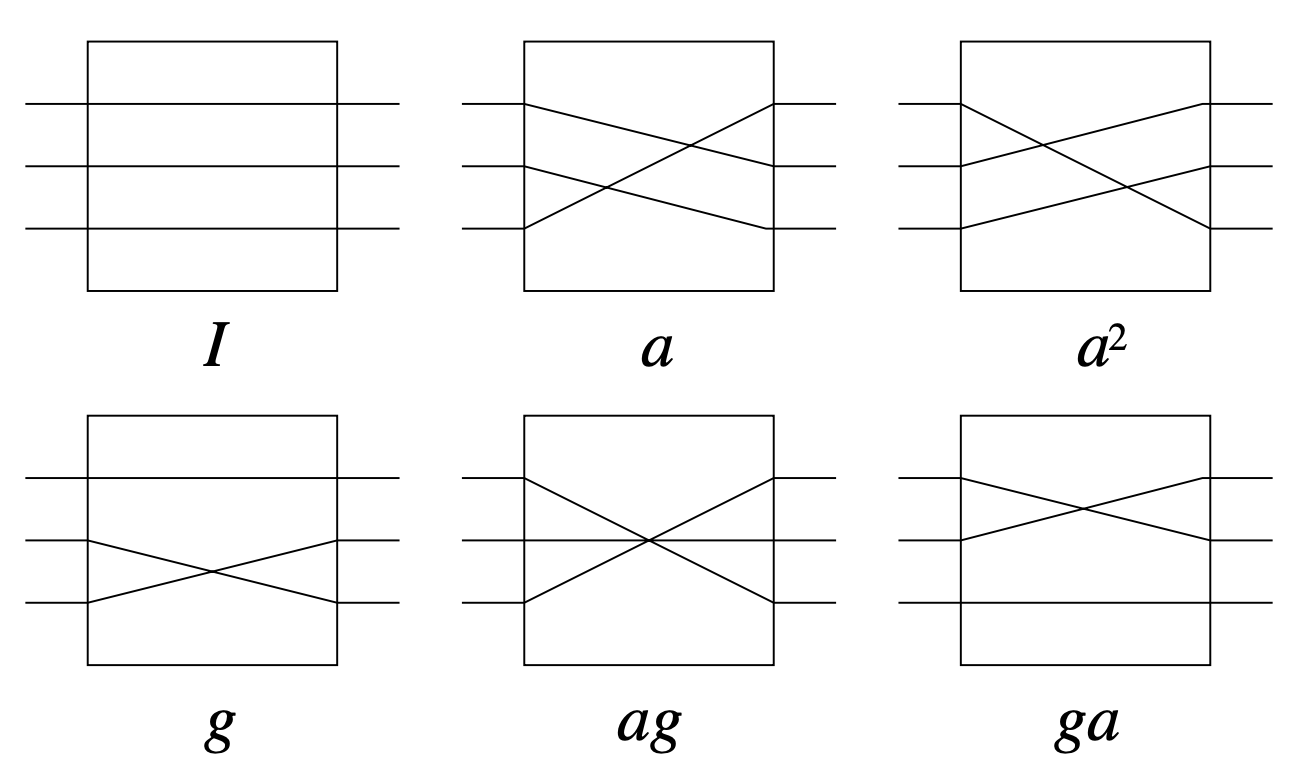}
    \caption{Elements of $D_3$}
    \label{f5}
\end{figure}

Since $gag = a^{-1}$, $a^3 = I$, and $g^2 = I$,  \[(ag)g = a(gg) = aI = a\] \[gag = a^{-1} = a^2\]
\[a^2g = (g^{-1}g)a^2g = g^{-1}(ga^2g) = g^{-1}a^{-2} = ga\]

The complete group map of $D_3$ is shown in Figure \ref{f6}.

\begin{figure}[H]
    \centering
    \includegraphics[width=6cm]{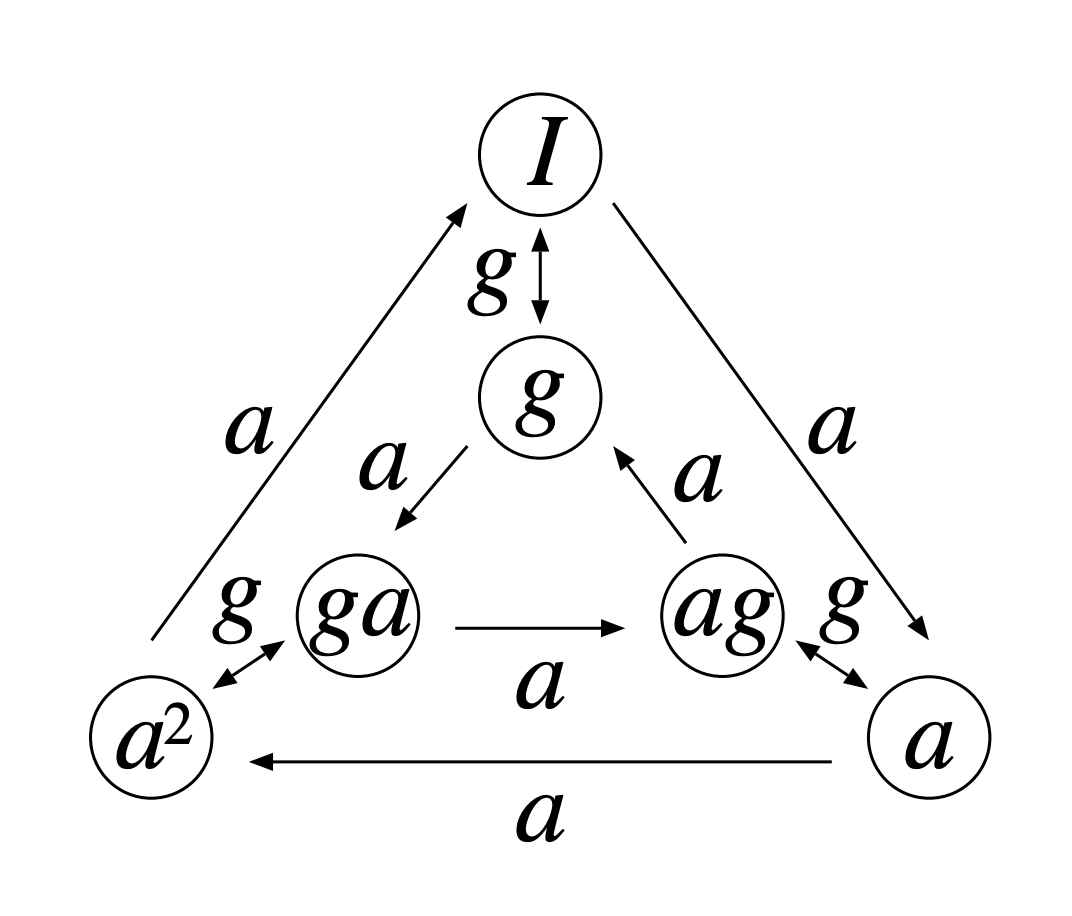}
    \caption{Group Map of $D_3$}
    \label{f6}
\end{figure}

Note that $D_3$ is non-Abelian as $ag \neq ga$. 

\subsubsection{Group $D_5$}

The group $D_5$ can be generated in a similar manner as $D_3$. $D_5 = \langle a,g: a^5 = g^2 = I, gag=a^{-1} \rangle $. Elements $a$ and $g$ are shown in Figure \ref{f7} where $a$ permutes the $n$th rail to the $n+1$th rail and $g$ swaps the $n$th rail with the $p-n$th rail when $n \neq 0$.

\begin{figure}[!ht]
    \centering
    \includegraphics[width=4.5cm]{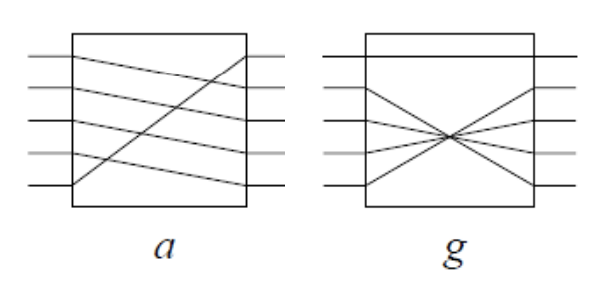}
    \caption{$D_5$ elements $a,g$}
    \label{f7}
\end{figure}

Elements $a$ and $g$ can be repeatedly composed to generate the complete dihedral group of order 10. The group map of $D_5$ is shown in Figure \ref{f8}.

\begin{figure}[!ht]
    \centering
    \includegraphics[width=4.5cm]{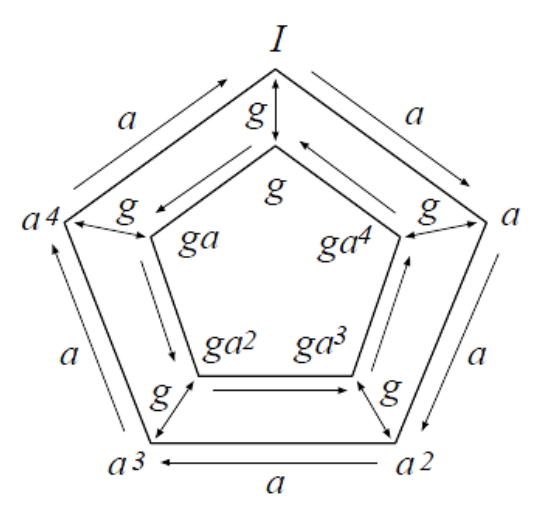}
    \caption{$D_5$ Group Map}
    \label{f8}
\end{figure}

Once again $gag= a^{-1}$ and $ga^{-1}g = a$.

\subsubsection{Dihedral Group of Order N}

In general, the group $D_n$ for arbitrary prime number $n$ can be generated using the elements $a$ and $g$ such that $a$ permutes the $n$th rail to the $n+1$th rail where the last rail maps to the top rail, and $g$ swaps the $i$th rail with the $p-i$th rail for all $i \neq 0$. The group map of $D_n$ is illustrated in Figure \ref{f9}.

\begin{figure}[!ht]
    \centering
    \scalebox{.85}{
    \includegraphics[width=6.5cm]{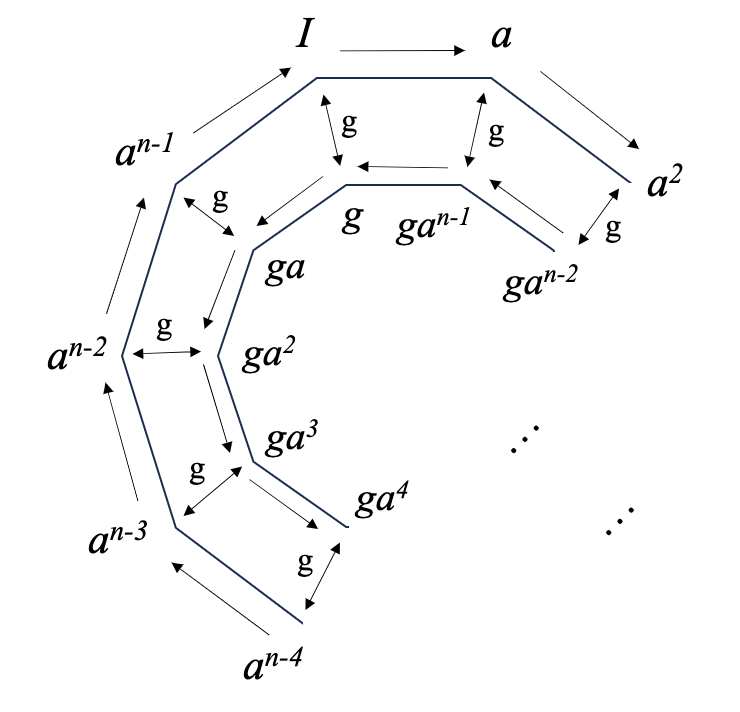}}
    \caption{$D_n$ Group Map}
    \label{f9}
\end{figure}

}

\section{Group Function Decomposition}

{\parindent0pt

\subsection{Single Input Variable}
Let $G$ be a group and $B = \{0,1\}$. Then, $F: B^n \rightarrow G$ is a group function. By Shannon's Expansion [\ref{f31}], a group function $F(x):B^n \rightarrow D_3$ decomposes as follows: 
\begin{equation}F(\hat{X},x_n) = F_a(\hat{X})g^{x_n}F_b(\hat{X})g^{x_n} \label{feq1} \tag{3.1} \end{equation} 
where $x_n$ is a two-valued input variable and   $F_a(\hat{X})$ and $F_b(\hat{X})$ denote group functions that do not depend on $x_n$ with $\hat{X}=(x_1, x_2, \ldots x_{m-1}) \in X^{m-1}$. A proof can be found in [\ref{f10}]. This decomposition is similar to Shannon’s expansion for a classical binary logic function and is essential in the design of canonical cascades. Figure \ref{f10} shows the canonical cascade function of 3.1:

\begin{figure}[!ht]
    \centering
    \includegraphics[width=8cm]{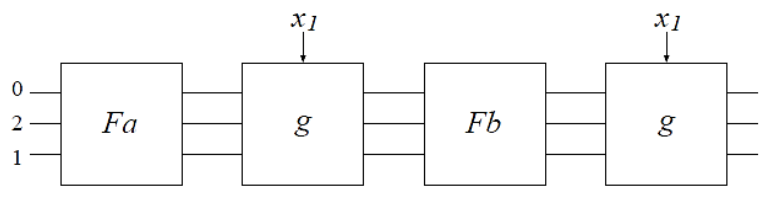}
    \caption{One Variable Canonical Cascade}
    \label{f10}
\end{figure}

The decomposition indicates that element $g$ needs to be modified so that it can be controlled with a binary input variable. When $x = 1$, $g^x$ swaps the bottom two lines and when $x=0$, $g^x$ performs the identity permutation. Figure \ref{f11} shows a representation of $g^x$. Note that the $g^x$ element resembles a reversible Controlled SWAP or Fredkin gate.

\begin{figure}[H]
    \centering
    \includegraphics[width=6cm]{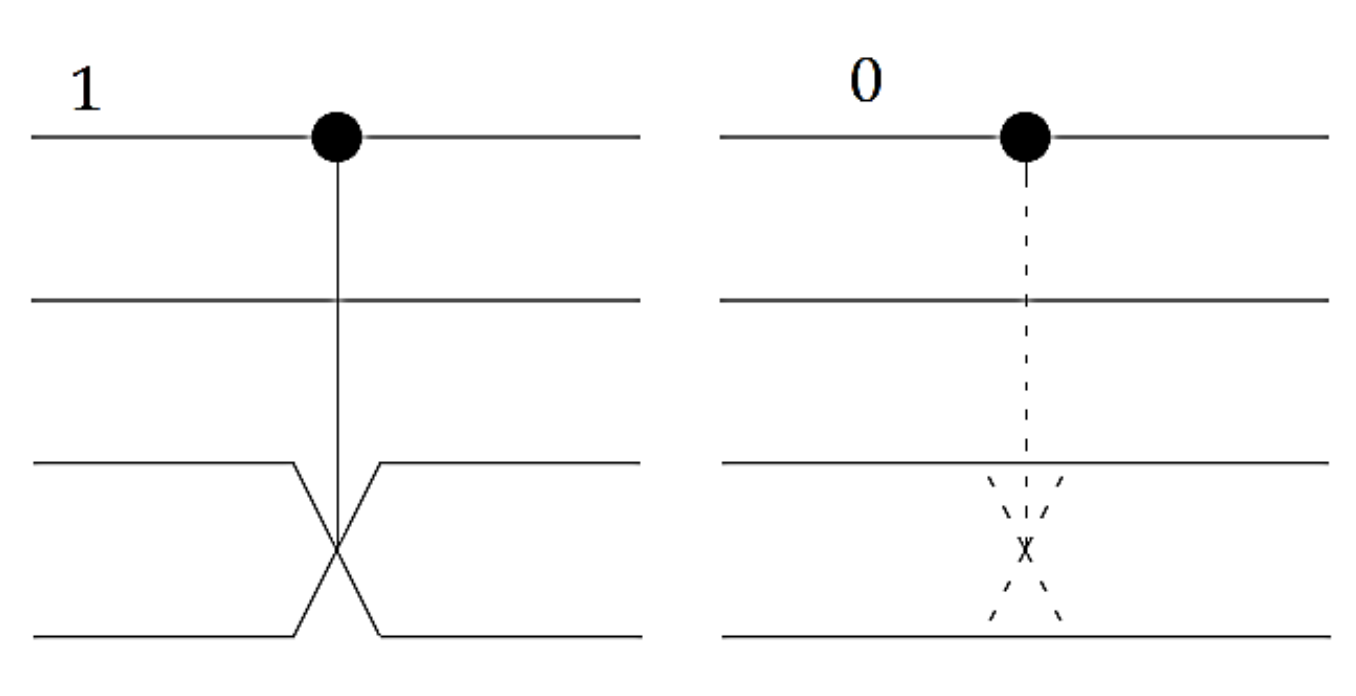}
    \caption{Element $g$ modified to $g^{x_n}$ so it can be controlled}
    \label{f11}
\end{figure}

Since $F_a(\hat{X})$ and $F_b(\hat{X})$ denote arbitrary group functions that do not depend on $x_n$ with $\hat{X}=(x_1, x_2, \ldots x_{m-1}) \in X^{m-1}$, we can define them to be functions of the group element  $a$. More specifically, assume $F_a(\hat{X}) = a^{f_a(\hat{X})}$ and $F_b(\hat{X}) = a^{f_b(\hat{X})}$ where $a$ is the shift element from our group.

From 3.1 we get \[F(\hat{X},x_n) = F_a(\hat{X})g^{x_n}F_b(\hat{X})g^{x_n} = a^{f_a(\hat{X})}g^{x_n}a^{f_b(\hat{X})}g^{x_n} \]
or equivalently,
\[a^{f(x_n)} = a^{f_a(\hat{X})}g^{x_n}a^{f_b(\hat{X})}g^{x_n}\label{feq:2} \tag{3.2}\]

Therefore, \[F(\hat{X},0) = a^{f_a(\hat{X})}g^{0}a^{f_b(\hat{X})}g^{0}\] \[F(\hat{X},1) = a^{f_a(\hat{X})}g^{1}a^{f_b(\hat{X})}g^{1}\]

Since $g^0 = I$ and $gag = a^{-1}$, we get 
\[F(\hat{X},0) = a^{f_a(\hat{X}) + f_b(\hat{X})}\] \[F(\hat{X},1) = a^{f_a(\hat{X}) - f_b(\hat{X})}\]

For convenience, define $F(\hat{X},k) = a^{f(\hat{X},k)}$ to express the above equations in terms of their exponents only:

\[f(\hat{X},0) = {f_a(\hat{X}) + f_b(\hat{X})}\] \[f(\hat{X},1) = {f_a(\hat{X}) - f_b(\hat{X})}\]

Converting the above expressions into matrix form we get, 
\[\begin{bmatrix}
f(\hat{X},0) \\
f(\hat{X},1) 
\end{bmatrix} = \begin{bmatrix}
+1 & +1 \\
+1 & -1
\end{bmatrix} \begin{bmatrix}
f_a(\hat{X}) \\
f_b(\hat{X}) 
\end{bmatrix}
\]

\[\Rightarrow \vec{F} = W_1 \vec{w} \label{feq:3} \tag{3.3}\]
where $\vec{F}$ is the truth vector of the function and $W_1$ is the first Walsh Matrix. Walsh transform as well as matrix representation of this transform are presented in [8, 13, 21, 22, 57]. Note that $f_a(\hat{X})$ and $f_b(\hat{X})$ represent the exponents of $a$ and literally describe the canonical form of the circuit cascade. If we let $\vec{w} = \begin{bmatrix}
f_a(\hat{X}) \\
f_b(\hat{X}) 
\end{bmatrix} = \begin{bmatrix}
w_a \\
w_b 
\end{bmatrix}$, then $w_a$ and  $w_b$ are the exponents of $a$ in our cascade. Hence, the canonical cascade can be found by solving equation 3.3 for $\vec{w}$ where $\vec{w}$ is the Walsh Spectrum of $\vec{F}$. Multiplying both sides of equation 3.3 by $W_1^{-1}$ we get, \[\vec{w} = (W_1^{-1}\vec{F}) \label{feq:4} \tag{3.4}\]

\subsubsection{Examples}

\textbf{Example 1: }
Consider the function $f(x) = x+1$, where $+$ denotes arithmetic addition. We will use the group $D_3$ to create our cascade. The truth vector for this function is $\vec{F} = \begin{bmatrix}
f(0) \\
f(1) 
\end{bmatrix} = \begin{bmatrix}
1 \\
2 
\end{bmatrix}$. Hence, from equation 3.4 we have $\vec{w} = (W_1^{-1})\vec{F} = (W_1^{-1})\begin{bmatrix}
1 \\
2 
\end{bmatrix}$. 

We now compute the inverse of the first Walsh Matrix. Since $W_1 = \begin{bmatrix}
+1 & +1 \\
+1 & -1
\end{bmatrix}$, \[W_1^2 = \begin{bmatrix}
2 & 0 \\
0 & 2
\end{bmatrix} = 2\begin{bmatrix}
1 & 0 \\
0 & 1
\end{bmatrix} = 2I = 2W_1W_1^{-1}\] Thus, $W_1^{-1} \equiv -W_1$ (mod 3). Therefore, \[\vec{w} = -W_1\vec{F} = \begin{bmatrix}
-3 & 1\\ 
\end{bmatrix} ^ T \equiv \begin{bmatrix}
0 & 1\\ 
\end{bmatrix} ^ T \text{ (mod 3)}\] 

From equation 3.2 we have $a^{f(x)} = a^{w_a}g^{x}a^{w_b}g^{x}$. Replacing $w_a$ and $w_b$ with $0$ and $1$ respectively, we get 
\[a^{f(x)} = a^{0}g^{x}a^{1}g^{x} = g^{x}a^1g^{x}\]

The corresponding cascade diagram with the internal structure of the gates is shown in Figure \ref{f12}.

\begin{figure}[!ht]
    \centering
    \includegraphics[width=7cm]{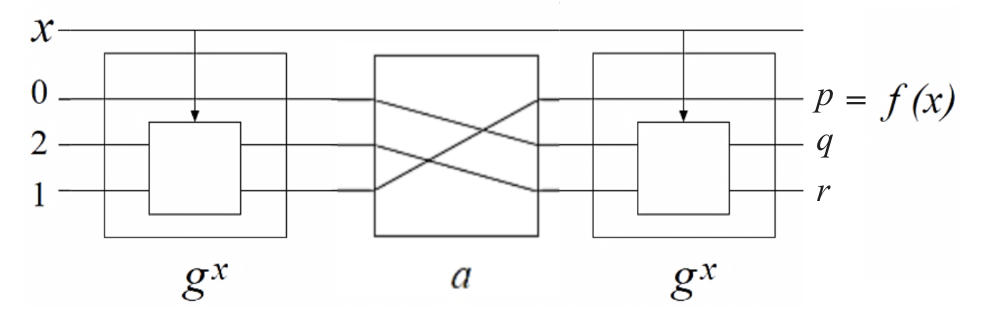}
    \caption{Canonical Cascade for $f(x) = x+1$}
    \label{f12}
\end{figure}

Note that only the top target line is used for the function output and the last $g$ gate does not affect that line. Hence we can remove the last $g$ gate without changing our output. Hence, the final expression for this function is $g^{x}a^{-1}$. The reduced cascade is shown in Figure \ref{f13}.

\begin{figure}[!ht]
    \centering
    \includegraphics[width=6cm]{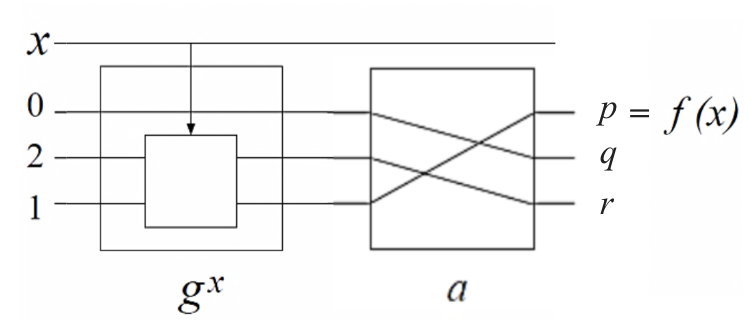}
    \caption{Reduced canonical cascade for $f(x) = x+1$ without final $g$ gate}
    \label{f13}
\end{figure}

We now create the circuit for the expression $g^{x}a^{-1}$. The truth table for the expression $g^{x}a^{-1}$ is shown in Figure \ref{f14}.

\begin{figure}[!h]
    \centering
    \begin{tabular}{|c|c|c|c|}
    \textbf{$x$} & \textbf{$p$} & \textbf{$q$} & \textbf{$r$}\\

    \hline
    0 & 1 & 0 & 2\\
    1 & 2 & 0 & 1
    \end{tabular}
    \caption{$f(x) = x+1$ cascade truth table}
    \label{f14}
\end{figure}

When $x = 0$, $0$ and $2$ are swapped and then $2$ and $1$ are swapped. When $x = 1$, $2$ and $0$ are swapped. The corresponding circuit diagram is shown in Figure \ref{f15}.

\begin{figure}[!ht]
    \centering
    \begin{quantikz}
    & \lstick{$x$} & \ctrl{2} & \targ{} & \ctrl{2} & \ctrl{3} & \qw \\
    & \lstick{0} & \swap{1} & \qw & \swap{1} & \swap{2} & \qw \rstick{$p = f(x)$} \\
    & \lstick{2} & \swap{0} & \qw & \swap{0} & \qw & \qw \rstick{$q$} \\
    & \lstick{1} & \qw & \qw & \qw & \swap{0} & \qw \rstick{$r$}
    \end{quantikz}
    \caption{$f(x) = x+1$ circuit}
    \label{f15}
\end{figure}
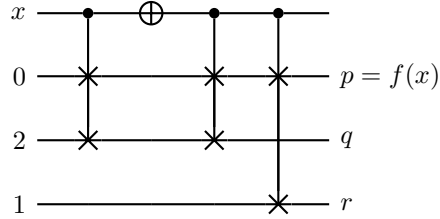

\textbf{Local Transformations:} We now apply local transformations to simplify our cascade. Note that we can replace the two controlled swap gates on the first and second target wires with a swap gate. This reduction is shown in Figure \ref{f16}.

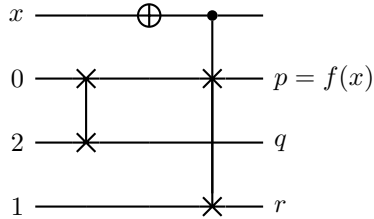
\begin{figure}[!ht]
    \centering
    \begin{quantikz}
    & \lstick{$x$} & \qw & \targ{} & \ctrl{3} & \qw \\
    & \lstick{0} & \swap{1} & \qw & \swap{2} & \qw \rstick{$p = f(x)$} \\
    & \lstick{2} & \swap{0} & \qw & \qw & \qw \rstick{$q$} \\
    & \lstick{1} & \qw & \qw & \swap{0} & \qw \rstick{$r$}
    \end{quantikz}
    \caption{$f(x) = x+1$ reduced circuit}
    \label{f16}
\end{figure}

However, we can remove the first swap gate and reorder the input signals from ``0, 2, 1" to ``2, 0, 1." Then, we can remove the inverter and move the output to the bottom wire. The final circuit for $f(x) = x+1$ is shown in Figure \ref{f17}.

\begin{figure}[!ht]
    \centering
    \begin{quantikz}
    & \lstick{$x$} & \ctrl{3} & \qw \\
    & \lstick{2}  & \swap{2} & \qw \rstick{$p$} \\
    & \lstick{0} & \qw & \qw \rstick{$q$} \\
    & \lstick{1} & \swap{0} & \qw \rstick{$r = f(x)$}
    \end{quantikz}
    \caption{$f(x) = x+1$ final circuit}
    \label{f17}
\end{figure}
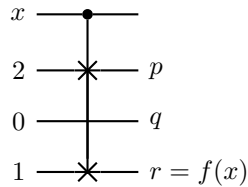

\subsection{Two Input Variables}

Expression 3.1 can be extended to two variables:
\[F(x_1,x_2) = F_a(x_2)g^{x_1}F_b(x_2)g^{x_1}\]
but in this case $F_a$ and $F_b$ are functions of one variable. Figure \ref{f18} shows the canonical cascade:

\begin{figure}[!ht]
    \centering
    \scalebox{0.8}{
    \includegraphics[width=9cm]{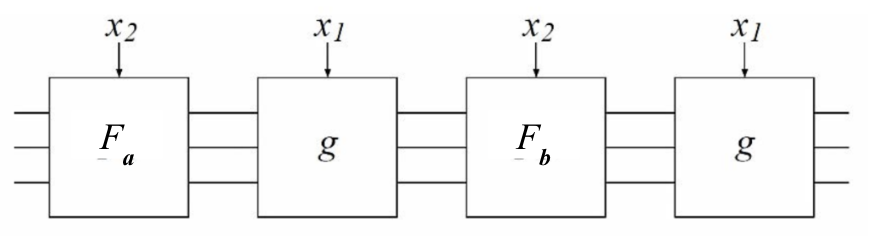}}
    \caption{Intermediate Cascade for two variable functions}
    \label{f18}
\end{figure}

From section 3.1 we found that a one variable function $F$ can be decomposed as $a^{w_a}g^{x}a^{w_b}g^x$. Since $F_a$ and $F_b$ are functions of one variable, we can replace them with $a^{w_a}g^{x_2}a^{w_b}g^{x_2}$ and $a^{w_c}g^{x_2}a^{w_d}g^{x_2}$ respectively. The canonical form for all functions with two input variables then becomes:

\[a^{f(x_1,x_2)} = ((a^{w_a}g^{x_2}a^{w_b}g^{x_2})g^{x_1})((a^{w_c}g^{x_2}a^{w_d}g^{x_2})g^{x_1})\]


The cascade has ten cells and is shown in Figure \ref{f19}.

\begin{figure}[!ht]
    \centering
    \includegraphics[width=10cm]{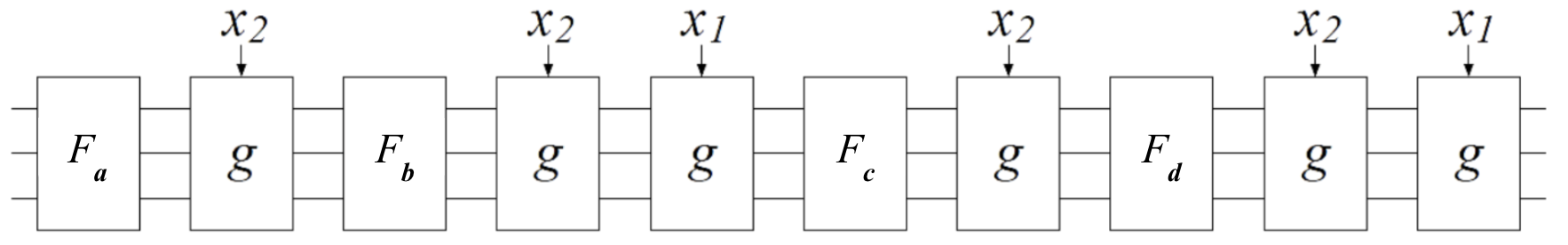}
    \caption{Cascade for two variable functions after decomposition}
    \label{f19}
\end{figure}

Next we will expand the Walsh matrix to two variables by assigning 0 and 1 to $x_1$ and $x_2$ for all four possible combinations and apply vectors to the exponents as was done in section 3.1 for the single variable case. After
applying the property of $gag = a^{-1}$ from Section 2.1, the following four equations are derived:


\[a^{f(0,0)} = ((a^{w_a}g^{0}a^{w_b}g^{0})g^{0})((a^{0}g^{x_2}a^{0}g^{x_2})g^{0}) = a^{w_a+w_b+w_c+w_d}\]
\[\Rightarrow f(0,0) = w_a+w_b+w_c+w_d\]

\[a^{f(0,1)} = ((a^{w_a}g^{1}a^{w_b}g^{1})g^{0})((a^{w_c}g^{1}a^{w_d}g^{1})g^{0}) = (a^{w_a}a^{-w_b})(a^{w_c}a^{-w_d}) = a^{w_a-w_b+w_c-w_d}\]
\[\Rightarrow f(0,1) = w_a-w_b+w_c-w_d\]

\[a^{f(1,0)} = ((a^{w_a}g^{0}a^{w_b}g^{0})g^{1})((a^{w_c}g^{0}a^{w_d}g^{0})g^{1}) = (a^{w_a+w_b})g^{1}(a^{w_c+w_d})g^{1}= a^{w_a+w_b}a^{-(w_c+w_d)}=a^{w_a+w_b-w_c-w_d}\]
\[\Rightarrow f(1,0) = w_a+w_b-w_c-w_d\]

\[a^{f(1,1)} = ((a^{w_a}g^{1}a^{w_b}g^{1})g^{1})((a^{w_c}g^{1}a^{w_d}g^{1})g^{1}) = ((a^{w_a}a^{-w_b})g^{1})((a^{w_c}a^{-w_d})g^{1}) = (a^{w_a-w_b})(a^{-(w_c-w_d)}) = a^{w_a-w_b-w_c+w_d}\]
\[\Rightarrow f(1,1) = w_a-w_b-w_c+w_d\]

Putting the above equations in matrix form, we obtain 

\[\begin{bmatrix}
    f(0,0) \\ f(0,1) \\ f(1,0) \\ f(1,1)
\end{bmatrix} = \begin{bmatrix}
    1 & 1 & 1 & 1 \\
    1 & -1 & 1 & -1 \\
    1 & 1 & -1 & -1 \\
    1 & -1 & -1 & 1 
\end{bmatrix}\begin{bmatrix}
    w_a \\ w_b \\ w_c \\ w_d
\end{bmatrix}\]

Thus, just as in the one variable case, the elements of the $w$ vector in the canonical cascade can be found using the equation 
\[\vec{F} = W_2 \vec{w}\]
\[\vec{w} = (W_2^{-1}\vec{F}) \label{feq:5} \tag{3.5}\]
where $W_2$ is the second Walsh matrix. This will be demonstrated in the next section.

\subsubsection{Examples}

\textbf{Example 2:} Consider the function $f(x_1,x_2) = x_1 \oplus x_2$, a simple binary XOR function. We will use $D_3$ to create our cascade. The truth vector for this function is $\vec{F} = \begin{bmatrix} 0 & 1 & 1 & 0\end{bmatrix}^T$. Hence, from equation 3.5 we have \[\vec{w} = (W_2^{-1})\vec{F} = (W_2^{-1})\begin{bmatrix}
0 \\ 1 \\ 1 \\ 0 \end{bmatrix}\]

We now compute the inverse of the second Walsh Matrix. 


\[W_2^2 = \begin{bmatrix}
4 & 0 & 0 & 0 \\
0 & 4 & 0 & 0 \\
0 & 0 & 4 & 0 \\
0 & 0 & 0 & 4 
\end{bmatrix} = 4\begin{bmatrix}
1 & 0 & 0 & 0 \\
0 & 1 & 0 & 0 \\
0 & 0 & 1 & 0 \\
0 & 0 & 0 & 1 
\end{bmatrix} = 4I_2 = 4W_2W_2^{-1}\]
\[\Rightarrow W_2^{-1} \equiv W_2  \text{ (mod 3)}\]

Thus, \[\vec{w} = (W_2^{-1})\vec{F} = W_2\begin{bmatrix}
0 \\ 1 \\ 1 \\ 0 \end{bmatrix} = \begin{bmatrix}
2 \\ 0 \\ 0 \\ -2 \end{bmatrix} \equiv \begin{bmatrix}
-1 \\ 0 \\ 0 \\1 \end{bmatrix} \text{ (mod 3)}\]

\[\vec{w} = \begin{bmatrix}
    w_a \\ w_b \\ w_c \\ w_d
\end{bmatrix} = \begin{bmatrix}
    -1 \\ 0 \\ 0 \\1
\end{bmatrix} \text{ (mod 3)}\]

The group decomposition expression for two variables in canonical form is
\[a^{f(x_1,x_2)} = a^{w_a}g^{x_2}a^{w_b}g^{x_2+x_1}a^{w_c}g^{x_2}a^{w_d}g^{x_2 + x_1}\]

After replacing the exponents for $w_a, w_b, w_c, \text{ and } w_d$ with the values obtained for $\vec{w},$ we have the following canonical cascade function:
\[a^{f(x_1,x_2)} = a^{-1}g^{x_2}a^{0}g^{x_2+x_1}a^{0}g^{x_2}a^{1}g^{x_2 + x_1} = a^{-1}g^{x_1+x_2}a^1g^{x_1+x_2}\]

The corresponding cascade is shown in Figure \ref{f20}.

\begin{figure}[!ht]
    \centering
    \includegraphics[width=7.5cm]{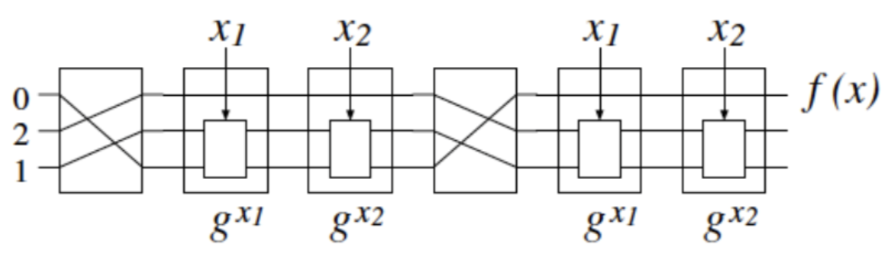}
    \caption{Cascade for $f(x_1,x_2) = x_1 \oplus x_2$}
    \label{f20}
\end{figure}

\textbf{Local Transformations:} We now apply local transformations to reduce our circuit. Notice that the last two $g$ gates can be removed as they do not affect the top target wire as shown in Figure \ref{f21}. 

\begin{figure}[!ht]
    \centering
    \includegraphics[width=6.5cm]{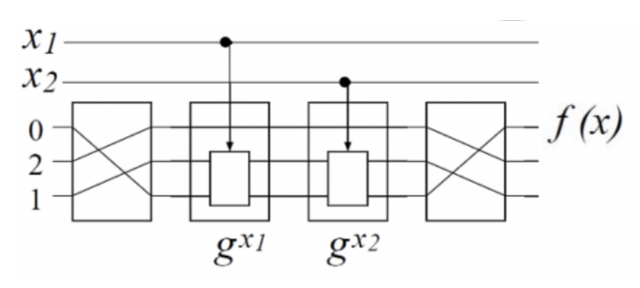}
    \caption{Reduced Cascade for $f(x_1,x_2) = x_1 \oplus x_2$}
    \label{f21}
\end{figure}

The last $a$ gate can be removed and the output can be moved to the bottom wire, and the first $a$ gate can be removed and the input signals can be rearranged. The final cascade for $f$ is shown in Figure \ref{f22}.

\begin{figure}[H]
    \centering
    \includegraphics[width=4cm]{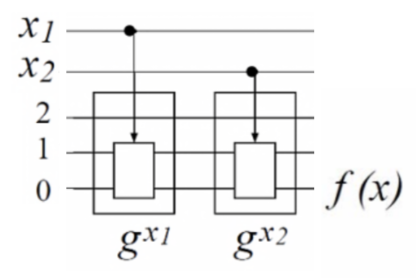}
    \caption{Final Cascade for $f(x_1,x_2) = x_1 \oplus x_2$}
    \label{f22}
\end{figure}

The quantum circuit for $f$ is shown in Figure \ref{f23}.

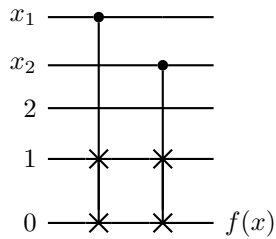
\begin{figure}[H]
    \centering
    \begin{quantikz}
    & \lstick{$x_1$} & \ctrl{4} & \qw & \qw \\
    & \lstick{$x_2$} & \qw & \ctrl{3} & \qw \\
    & \lstick{2} & \qw & \qw & \qw  \\
    & \lstick{1} & \swap{1} & \swap{1} & \qw  \\
    & \lstick{0} & \swap{0} & \swap{0} & \qw \rstick{$f(x)$}
    \end{quantikz}
    \caption{$f(x) = x_1 \oplus x_2$ circuit}
    \label{f23}
\end{figure}


\subsection{Three Input Variables}

Expression 3.1 can also be extended to three variables and after decomposition, the following canonical expression is derived:

\[a^{f(x_1,x_2,x_3)} = [[((a^{w_a}g^{x_3}a^{w_b}g^{x_3})g^{x_2})((a^{w_c}g^{x_3}a^{w_d}g^{x_3})g^{x_2})]g^{x_1}][[((a^{w_e}g^{x_3}a^{w_f}g^{x_3})g^{x_2})((a^{w_g}g^{x_3}a^{w_h}g^{x_3})g^{x_2})]g^{x_1}]\]

\[= a^{w_a}g^{x_3}a^{w_b}g^{x_2 + x_3}a^{w_c}g^{x_3}a^{w_d}g^{x_1+x_2 + x_3}a^{w_e}g^{x_3}a^{w_f}g^{x_2 + x_3}a^{w_g}g^{x_3}a^{w_h}g^{x_1 + x_2+x_3}\]

\subsubsection{Examples: }

\textbf{Example 3:}
Let $f(x_1, x_2, x_3) = x_1+x_2+x_3$, a binary input, ternary output adder using the group $D_3$. The truth table for $f$ is shown in Figure \ref{f24}:

\begin{figure}[!h]
    \centering
    \begin{tabular}{|c|c|c|c|}
    \textbf{$x_3$} & \textbf{$x_2$} & \textbf{$x_1$} & \textbf{$f(x_1, x_2, x_3)$} \\

    \hline
    0 & 0 & 0 & 0\\
    0 & 0 & 1 & 1\\
    0 & 1 & 0 & 1\\
    0 & 1 & 1 & 2\\
    1 & 0 & 0 & 1\\
    1 & 0 & 1 & 2\\
    1 & 1 & 0 & 2\\
    1 & 1 & 1 & 0
    \end{tabular}
    \caption{Modulo 3 Adder Truth Table}
    \label{f24}
\end{figure}

The truth vector for this function is $\vec{F} = [0, 1, 1, 2, 1, 2, 2, 0]^T$. The inverse Walsh matrix ${W_3}^{-1} \equiv -W_3$ (mod 3), so \[\vec{w} = ({W_3}^{-1})\vec{F} \equiv -W_3\vec{F} \text{ (mod 3)}\] \[= [0, 1, 1, 0, 1, 0, 0, 0]^T \text{ (mod 3)}\] Thus, \[f(x_1,x_2,x_3) = g^{x_3}a^1g^{x_2+x_3}a^1g^{x_1+x_2}a^1g^{x_1+x_2+x_3}\] Since the last $g$ is unused, we can remove it to obtain $f(x_1,x_2,x_3) = g^{x_3}a^1g^{x_2+x_3}a^1g^{x_1+x_2}a^1$. The cascade is shown in Figure \ref{f25}.

\begin{figure}[!h]
    \centering
    \includegraphics[width=9cm]{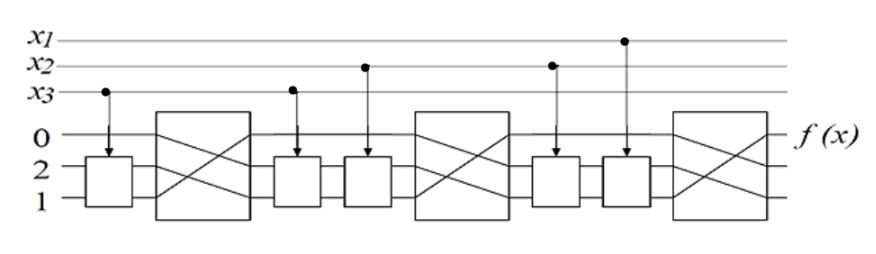}
    \caption{Canonical Cascade for the modulo three adder of binary arguments, $f(x_1, x_2, x_3) = x_1+x_2+x_3$}
    \label{f25}
\end{figure}

\textbf{Local Transformations:} The next phase will be to apply local transformations to simplify the cascade.

Consider the first three gates from Figure \ref{f25} as shown in Figure \ref{f26}. 

\begin{figure}[H]
    \centering
    \includegraphics[width=10cm]{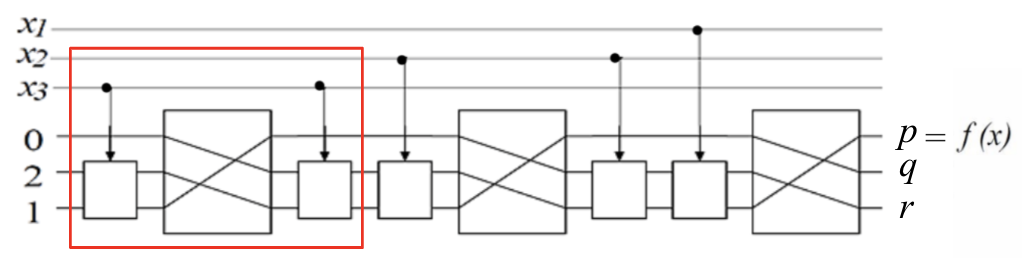}
    \caption{$g^{x_3}a^1g^{x_3}$}
    \label{f26}
\end{figure}

The internal structure of these three gates is shown in Figure \ref{f27}.

\begin{figure}[!ht]
    \centering
    \includegraphics[width=6cm]{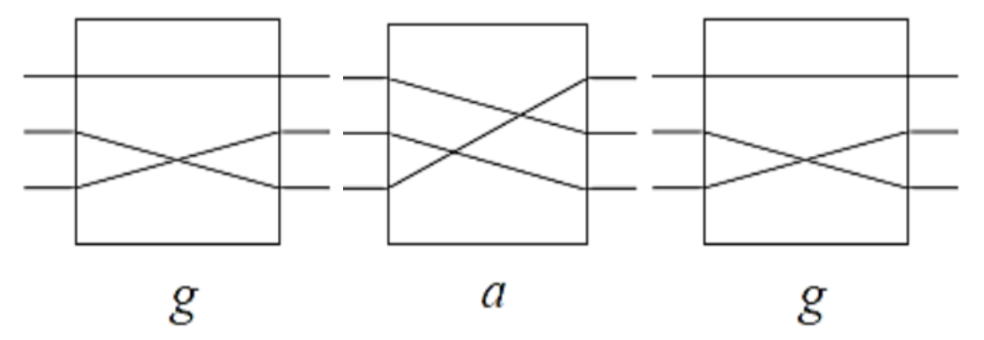}
    \caption{Internal Structure of $g^{x_3}a^1g^{x_3}$ I}
    \label{f27}
\end{figure}

Using $gag = a^{-1}$ when the control qubit is 1, the truth table for the function $g^{x_3}a^1g^{x_3}$ is shown in Figure \ref{f28}.

\begin{figure}[!h]
    \centering
    \begin{tabular}{|c|c|c|c|}
  \textbf{$x_3$} & \textbf{$p$} & \textbf{$q$} & \textbf{$r$} \\

    \hline
     0 & 1 & 0 & 2\\
     1 & 2 & 1 & 0
    \end{tabular}
    \caption{$g^{x_3}a^1g^{x_3}$ Truth Table}
    \label{f28}
\end{figure}

The binary input, ternary output quantum circuit with binary controlled, ternary Fredkin gates for $g^{x_3}a^1g^{x_3}$ is shown in Figure \ref{f29}.

\begin{figure}[H]
    \centering
    \scalebox{0.9}{
    \begin{quantikz}
    & \lstick{$x_3$} & \targ{} & \ctrl{2} & \ctrl{3} & \targ{} & \ctrl{3} & \ctrl{2} & \qw \\
    & \lstick{0} & \qw & \swap{1} & \swap{2} & \qw & \swap{2}  & \swap{1} & \qw \rstick{$p$} \\
    & \lstick{2} & \qw & \swap{0} & \qw & \qw & \qw & \swap{0} & \qw  \rstick{$q$} \\
    & \lstick{1} & \qw & \qw & \swap{0} & \qw & \swap{0} & \qw & \qw \rstick{$r$} 
    \end{quantikz}}
    \caption{$g^{x_3}a^1g^{x_3}$ circuit}
    \label{f29}
\end{figure}
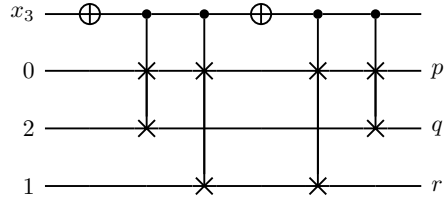

Note that the second and third controlled swap gates can be removed and replaced with a single swap gate and an inverter. The reduced circuit is shown in Figure \ref{f30}.

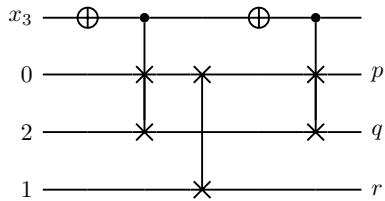
\begin{figure}[!ht]
    \centering
    \scalebox{0.9}{
    \begin{quantikz}
    & \lstick{$x_3$} & \targ{} & \ctrl{2} & \qw & \targ{} & \ctrl{2} & \qw \\
    & \lstick{0} & \qw & \swap{1} & \swap{2} & \qw & \swap{1} & \qw \rstick{$p$} \\
    & \lstick{2} & \qw & \swap{0} & \qw & \qw & \swap{0} & \qw  \rstick{$q$} \\
    & \lstick{1} & \qw & \qw & \swap{0} & \qw & \qw & \qw \rstick{$r$} 
    \end{quantikz}}
    \caption{Reduced $g^{x_3}a^1g^{x_3}$ circuit}
    \label{f30}
\end{figure}

Similarly, we can simplify the next three gates in our cascade, $g^{x_2}a^1g^{x_2}$ shown in Figure \ref{f31}.

\begin{figure}[!ht]
    \centering
    \includegraphics[width=7.3cm]{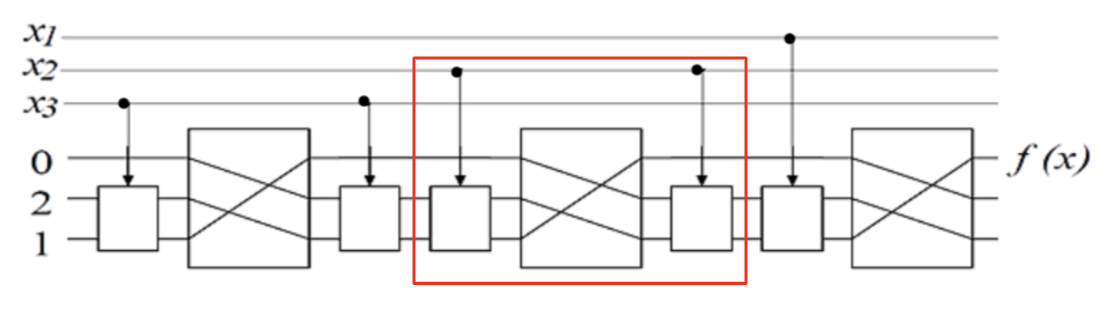}
    \caption{$g^{x_2}a^1g^{x_2}$}
    \label{f31}
\end{figure}

The circuit for these three gates is shown in Figure \ref{f32}.

\begin{figure}[!ht]
    \centering
    \begin{quantikz}
    & \lstick{$x_2$} & \targ{} & \ctrl{2} & \qw & \targ{} & \ctrl{2} & \qw \\
    & \lstick{0} & \qw & \swap{1} & \swap{2} & \qw & \swap{1} & \qw \rstick{$p$} \\
    & \lstick{2} & \qw & \swap{0} & \qw & \qw & \swap{0} & \qw  \rstick{$q$} \\
    & \lstick{1} & \qw & \qw & \swap{0} & \qw & \qw & \qw \rstick{$r$} 
    \end{quantikz}
    \caption{Reduced $g^{x_2}a^1g^{x_2}$ circuit}
    \label{f32}
\end{figure}
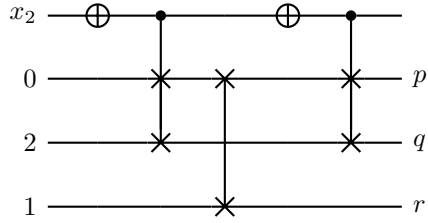

Now consider the last $a^1$ gate as shown in Figure \ref{f33}.

\begin{figure}[H]
    \centering
    \includegraphics[width=10cm]{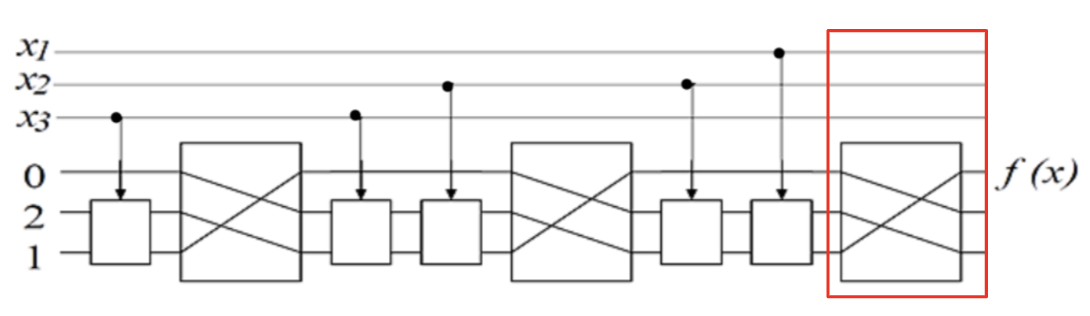}
    \caption{Last $a^1$ gate}
    \label{f33}
\end{figure}

It can be removed and we can move the output, $f(x)$, to the bottom wire instead as shown in Figure \ref{f34}.

\begin{figure}[!ht]
    \centering
    \includegraphics[width=8.5cm]{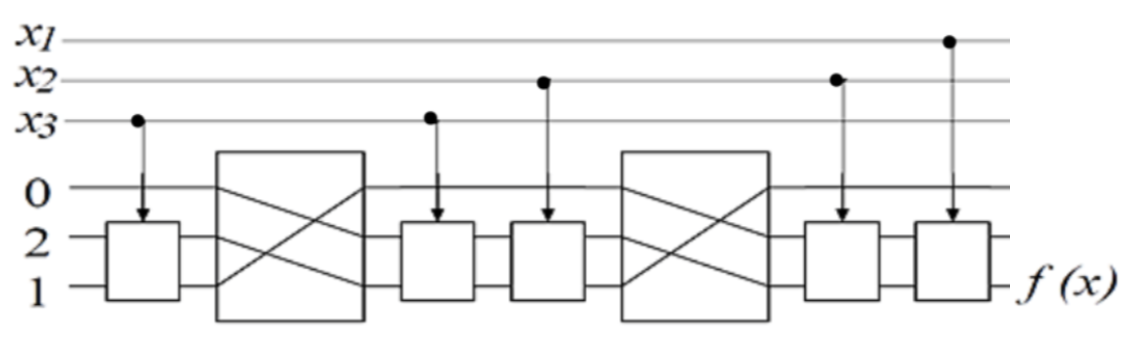}
    \caption{Simplified cascade with output on bottom wire}
    \label{f34}
\end{figure}

Combining our two reduced circuits and adding the circuit for $g^{x_1}$ we obtain the final circuit for $f(x_1, x_2, x_3) = x_1+x_2+x_3$ in Figure \ref{f35}.

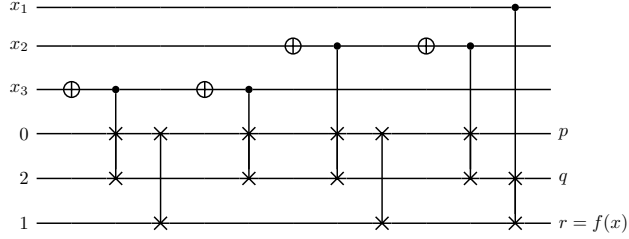
\begin{figure}[H]
    \centering
    \scalebox{0.7}{
    \begin{quantikz}
    & \lstick{$x_1$} & \qw & \qw & \qw & \qw & \qw & \qw & \qw & \qw & \qw & \qw & \ctrl{5} & \qw\\
    & \lstick{$x_2$} & \qw & \qw & \qw & \qw & \qw & \targ{} & \ctrl{3} & \qw & \targ{} & \ctrl{3} & \qw & \qw\\
    & \lstick{$x_3$} & \targ{} & \ctrl{2} & \qw & \targ{} & \ctrl{2} & \qw & \qw & \qw & \qw & \qw & \qw & \qw  \\
    & \lstick{0} & \qw & \swap{1} & \swap{2} & \qw & \swap{1}  & \qw & \swap{1} & \swap{2} & \qw & \swap{1} & \qw  & \qw \rstick{$p$} \\
    & \lstick{2} & \qw & \swap{0} & \qw & \qw & \swap{0} & \qw & \swap{0} & \qw & \qw & \swap{0} & \swap{1} & \qw  \rstick{$q$} \\
    & \lstick{1} & \qw & \qw & \swap{0} & \qw & \qw & \qw & \qw & \swap{0} & \qw & \qw & \swap{0} & \qw \rstick{$r=f(x)$}    
    \end{quantikz}}
    \caption{$f(x_1, x_2, x_3) = x_1+x_2+x_3 = g^{x_3}a^1g^{x_2+x_3}a^1g^{x_1+x_2}a^1$ circuit}
    \label{f35}
\end{figure}

The quantum layout for this circuit is shown in Figure \ref{f36}.
\begin{figure}[!ht]
    \centering
    \scalebox{.7}{
    \begin{tikzpicture}
\node[circle,draw, minimum size=0.7cm] (A) at  (0,0) {$x_2$};
\node[circle,draw, minimum size=0.7cm] (B) at  (0,1.4)  {2};
\node[circle,draw, minimum size=0.7cm] (C) at  (0,2.8) {0};
\node[circle,draw, minimum size=0.7cm] (D) at  (0,4.2) {$x_3$};

\node[circle,draw, minimum size=0.7cm] (E) at  (1.4,1.4) {$x_1$};
\node[circle,draw, minimum size=0.7cm] (F) at  (1.4,2.8) {1};

\draw (A) -- (B);
\draw (B) -- (C);
\draw (C) -- (D);
\draw (B) -- (E);
\draw (E) -- (F);
\draw (C) -- (F);
\end{tikzpicture}}
\caption{Quantum Layout for $f(x_1,x_2,x_3) = x_1+x_2+x_3$}
\label{f36}
\end{figure}
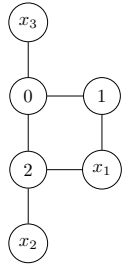

\subsection{Proof of Odd Valued Output}

We now prove that the group we use must have an odd valued output. From equation 3.4, in order to determine the Walsh Spectrum of our function, we need to compute the inverse of the Walsh matrix modulo $n$ where $n$ is our valued output. Since $W_i^2 = 2^iI$ where $I$ is the identity matrix, we can write \[W_i^2 = 2^iW_iW_i^{-1}\] Thus, $W_i^{-1} \equiv \frac{1}{2^i}W_i \text{ (mod }n\text{)}$. It remains to find the inverse of $2^i$ modulo $n$. However, a number, $k$, only has a multiplicative inverse modulo $n$ if $k$ and $n$ are relatively prime. Hence, in order for there to exist a multiplicative inverse of $2^i$ modulo $n$, $n$ must be odd.

}

\section{Local Transformations}

The local transformations applied in our methodology are as follows:
\begin{enumerate}
    \item Remove any $g$ gates at the end of the cascade as they do not affect the output wire.

    \begin{figure}[H]
        \centering
         \includegraphics[width=2.7cm]{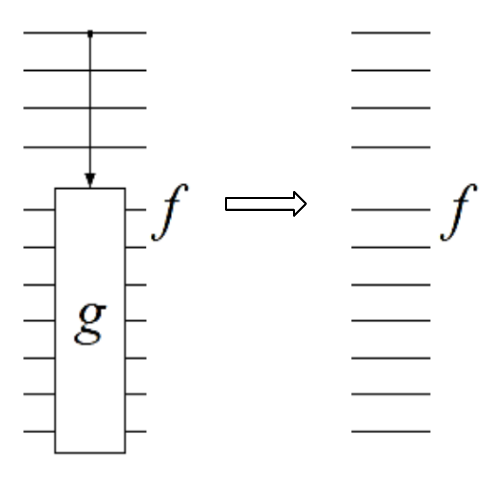}
        \caption{Removing $g$ gates at the end of the cascade}
        \label{f37}
    \end{figure}

    \item Remove any $a$ gates at the start of the cascade and reorder the input signals.

    \begin{figure}[!ht]
        \centering
        \includegraphics[width=3.1cm]{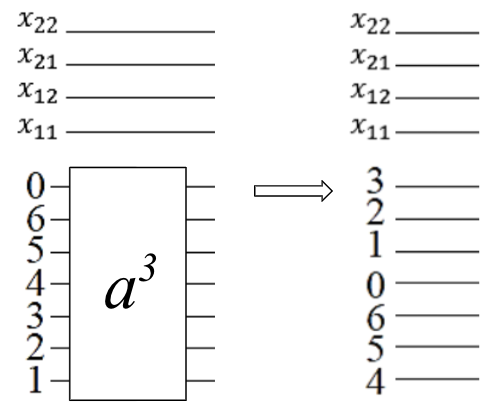}
        \caption{Reordering the input wires}
        \label{f38}
    \end{figure}

    \item Remove any $a$ gates at the end of the cascade and move the output to a different wire.

    \begin{figure}[!ht]
        \centering
        \includegraphics[width=3.7cm]{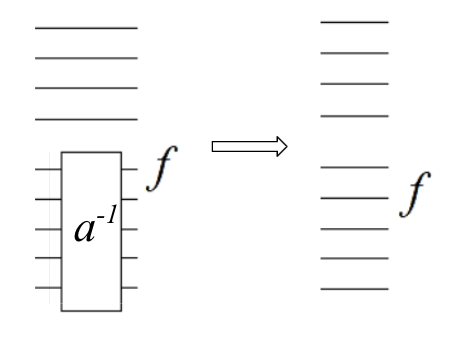}
        \caption{Changing the output wire}
        \label{f39}
    \end{figure}

    \item Replace any consecutive $g^xa^kg^x$ cells with $a^{-k}$, a NOT gate, and $a^k$. ($g^xa^kg^x = a^{-k}$ when $x = 1$ and $g^xa^kg^x = a^{k}$ when $x = 0$).

    \begin{figure}[H]
        \centering
        \includegraphics[width=7.8cm]{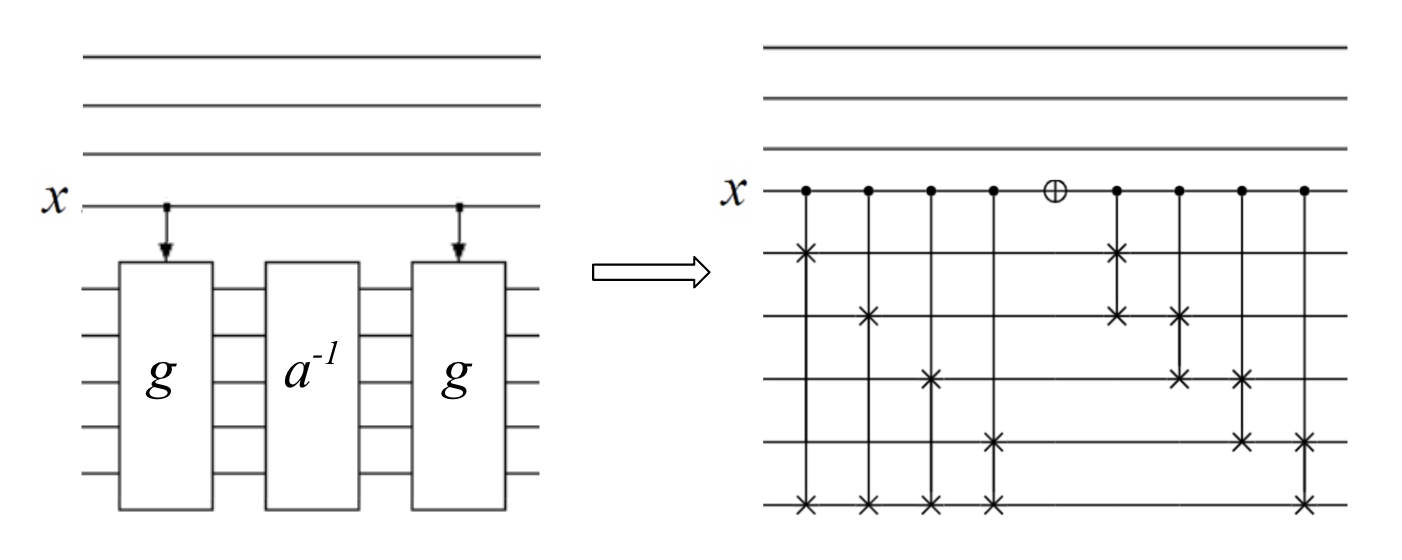}
        \caption{Reduction of consecutive $g^xa^{k}g^x$ cells with $k = -1$}
        \label{f40}
    \end{figure}

    \item Replace any consecutive $a^kg^xa^kg^x$ cells with a NOT gate and $a^{2k}$. ($a^kg^xa^kg^x = a^ka^{-k} = I$ when $x = 1$ and $a^kg^xa^kg^x = a^{2k}$ when $x = 0$).

    \begin{figure}[!ht]
        \centering
        \includegraphics[width=7.9cm]{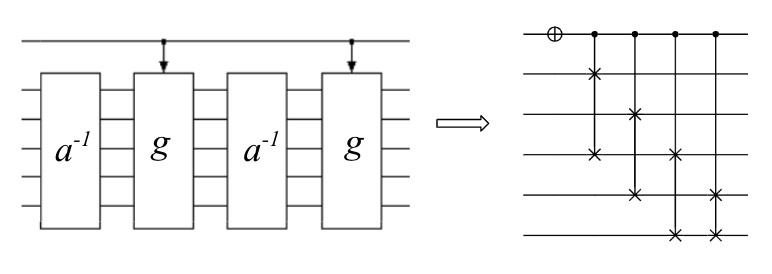}
        \caption{Reduction of consecutive $a^kg^xa^kg^x$ cells with $k = -1$}
        \label{f41}
    \end{figure}

    \item Replace any consecutive Controlled SWAP gate, a NOT gate, and another Controlled SWAP gate with a single swap gate

    \begin{figure}[H]
        \centering
        \includegraphics[width=4cm]{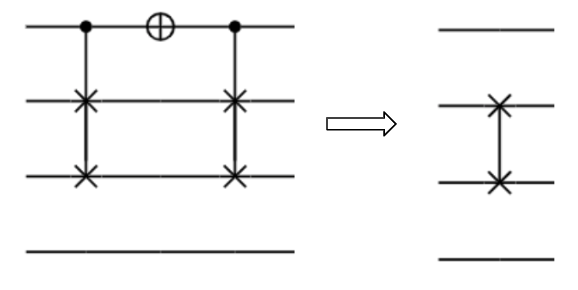}
        \caption{Reduction of consecutive CSWAP, NOT, CSWAP gates}
        \label{f42}
    \end{figure}

    \item Remove any SWAP gates at the start of the circuit and reorder the inputs of the wires

        \begin{figure}[!ht]
            \centering
            \includegraphics[width=3.6cm]{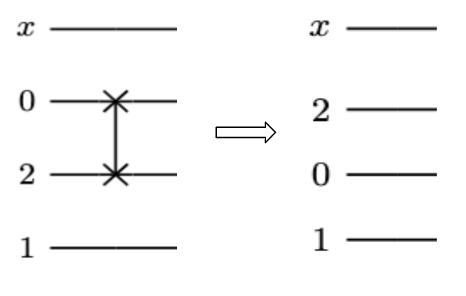}
            \caption{Removing swap gates at the beginning of the cascade}
            \label{f43}
        \end{figure}
\end{enumerate}

\section{Examples for 5 and 7 Valued Outputs}

{\parindent0pt

\textbf{Example 4:} Let $f(x_{22}, x_{12}, x_{21}, x_{11}) = 2(x_{22} + x_{21}) + (x_{12} + x_{11})$, a two-bit, four binary input variable, modulo 7 output adder. We will use the group $D_7$ to create the cascade. The truth table for $f$ is shown in Figure \ref{f44}.

\begin{figure}[!h]
    \centering
    \begin{tabular}{|c|c|c|c|c|}
    \textbf{$x_{22}$} & \textbf{$x_{12}$} & \textbf{$x_{21}$} & \textbf{$x_{11}$} & \textbf{$\vec{F}$} \\

    \hline
    0 & 0 & 0 & 0 & 0\\
    0 & 0 & 0 & 1 & 1\\
    0 & 0 & 1 & 0 & 2\\
    0 & 0 & 1 & 1 & 3\\

    0 & 1 & 0 & 0 & 1\\
    0 & 1 & 0 & 1 & 2\\
    0 & 1 & 1 & 0 & 3\\
    0 & 1 & 1 & 1 & 4\\

    1 & 0 & 0 & 0 & 2\\
    1 & 0 & 0 & 1 & 3\\
    1 & 0 & 1 & 0 & 4\\
    1 & 0 & 1 & 1 & 5\\

    1 & 1 & 0 & 0 & 3\\
    1 & 1 & 0 & 1 & 4\\
    1 & 1 & 1 & 0 & 5\\
    1 & 1 & 1 & 1 & 6
    \end{tabular}
    \caption{Two-Bit Modulo 7 Adder Truth Table}
    \label{f44}
\end{figure}

The truth vector is $\vec{F} = [0,1,2,3,1,2,3,4,2,3,4,5,3,4,5,6]^T$. Since ${W_4}^2 = 2^4I_4 \equiv 2I_4$ (mod 7),  ${W_4}^{-1} \equiv \frac{1}{2}W_4$ (mod 7). Thus, \[\vec{w} = ({W_4}^{-1})\vec{F} \equiv \frac{1}{2} W_4\vec{F} \text{ (mod 7)}\] \[= [3, 3, -1, 0, 3, 0, 0, 0, -1, 0, 0, 0, 0, 0, 0, 0]^T \text{ (mod 7)}\]

After simplification, the canonical cascade is the following: \[f = a^3g^{x_{11}}a^3g^{x_{11}}g^{x_{21}}a^{-1}g^{x_{21}}g^{x_{12}}a^3g^{x_{21}}g^{x_{22}}a^{-1}\]

The canonical cascade circuit diagram for $f$ is shown in Figure \ref{f45}.

\begin{figure}[!ht]
    \centering
    \includegraphics[width=8cm]{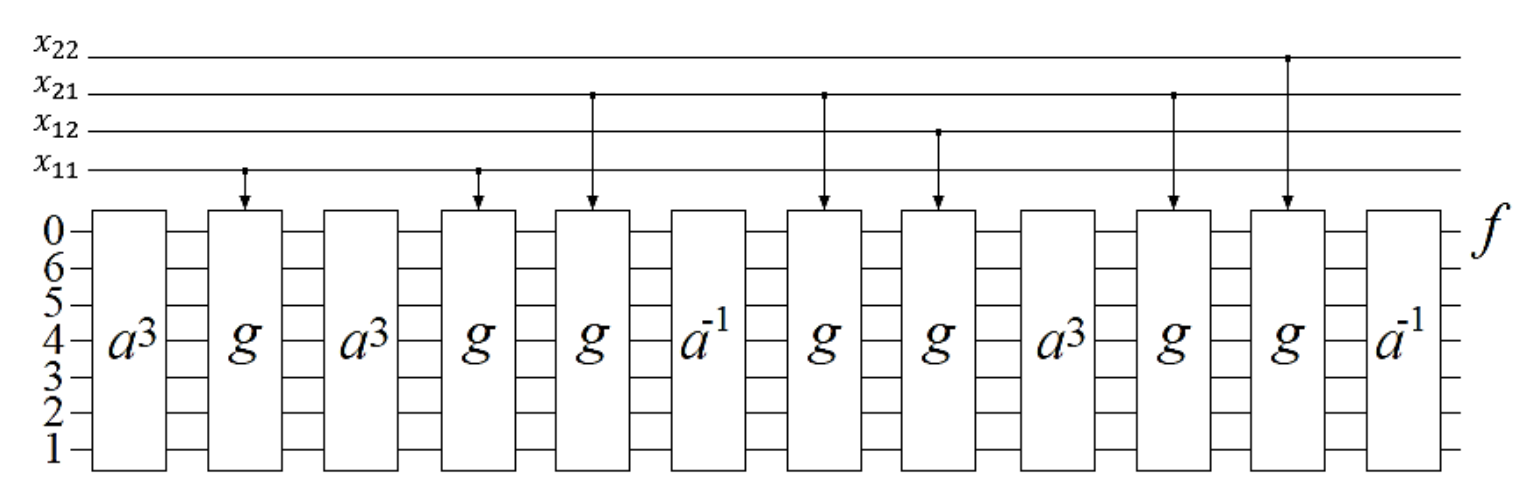}
    \caption{Reduced Canonical Cascade for two bit modulo 7 adder, $f = 2(x_{22} + x_{21}) + (x_{12} + x_{11})$}
    \label{f45}
\end{figure}

The internal structure of these gates is shown in Figure \ref{f46}:

\begin{figure}[H]
    \centering
    \includegraphics[width=2.8cm]{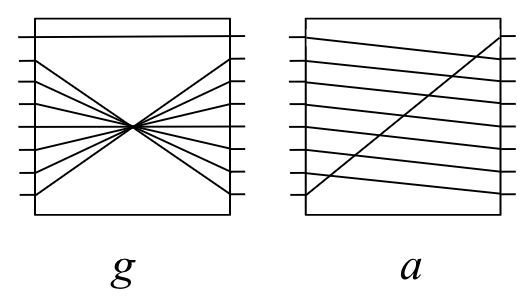}
    \caption{Internal Structure of $g$ and $a$ gates}
    \label{f46}
\end{figure}

\textbf{Local Transformations:} The next phase will be to apply local transformations to simplify the cascade.

Consider the first four cells shown in Figure \ref{f47}. 

\begin{figure}[!ht]
    \centering
    \includegraphics[width=8cm]{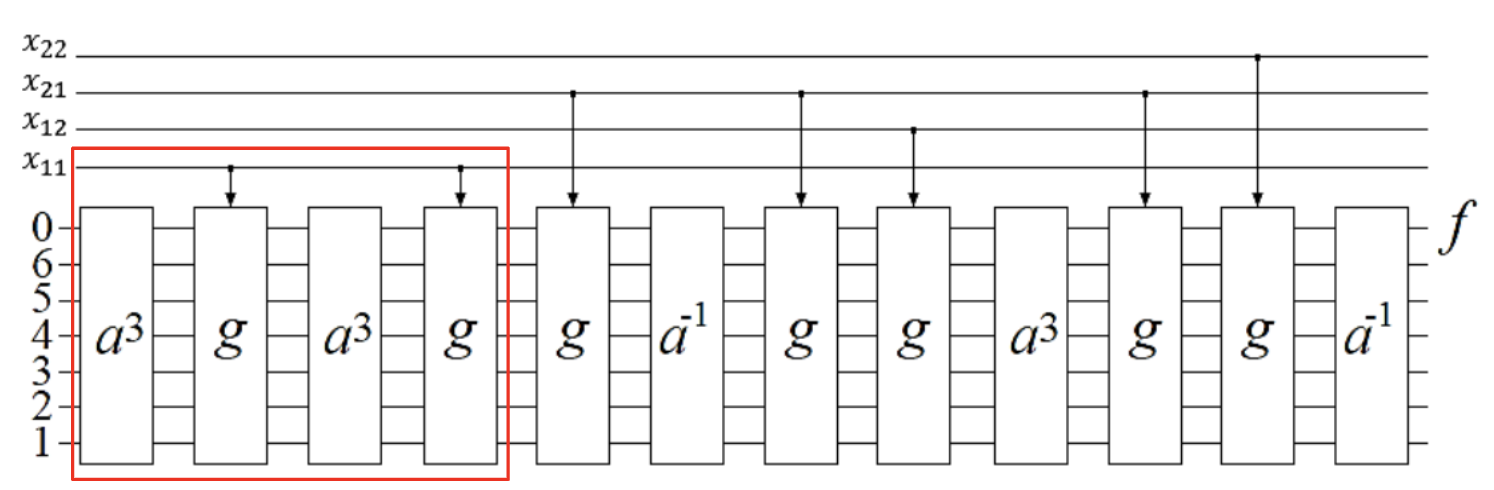}
    \caption{$a^3g^{x_{11}}a^3g^{x_{11}}$}
    \label{f47}
\end{figure}

Since $gag = a^{-1}$ when the control qubit is 1, when $x_{11} = 1$, the first four gates evaluate to $a^3a^{-3} = I$. When the control qubit is 0, the first four gates evaluate to $a^3a^3 = a^{-1}$. The circuit for $a^3g^{x_{11}}a^3g^{x_{11}}$ is shown in Figure \ref{f48}.

\begin{figure}[H]
    \centering
    \scalebox{.6}{
    \begin{quantikz}
    & \targ{} & \ctrl{2} & \ctrl{3} & \ctrl{4} & \ctrl{5} & \ctrl{6} & \ctrl{7} & \qw \\
    & \qw & \swap{1} & \qw & \qw  & \qw & \qw & \qw & \qw \\
    & \qw & \swap{0} & \swap{1} & \qw & \qw & \qw & \qw & \qw \\
    & \qw & \qw & \swap{0} & \swap{1} & \qw & \qw & \qw & \qw \\
    & \qw & \qw & \qw & \swap{0} & \swap{1} & \qw & \qw & \qw \\
    & \qw & \qw & \qw & \qw & \swap{0} & \swap{1} & \qw & \qw \\
    & \qw & \qw & \qw & \qw & \qw & \swap{0} & \swap{1} & \qw \\
    & \qw & \qw & \qw & \qw & \qw & \qw & \swap{0} & \qw \\
    \end{quantikz}}
    \caption{$a^3g^{x_{11}}a^{3}g^{x_{11}}$ circuit}
    \label{f48}
\end{figure}
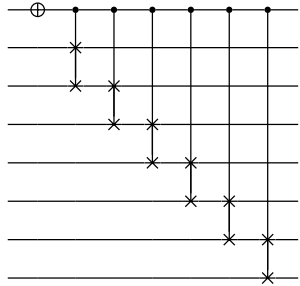

Similarly, we can simplify the three gates shown in Figure \ref{f49}. 

\begin{figure}[!h]
    \centering
    \includegraphics[width=9cm]{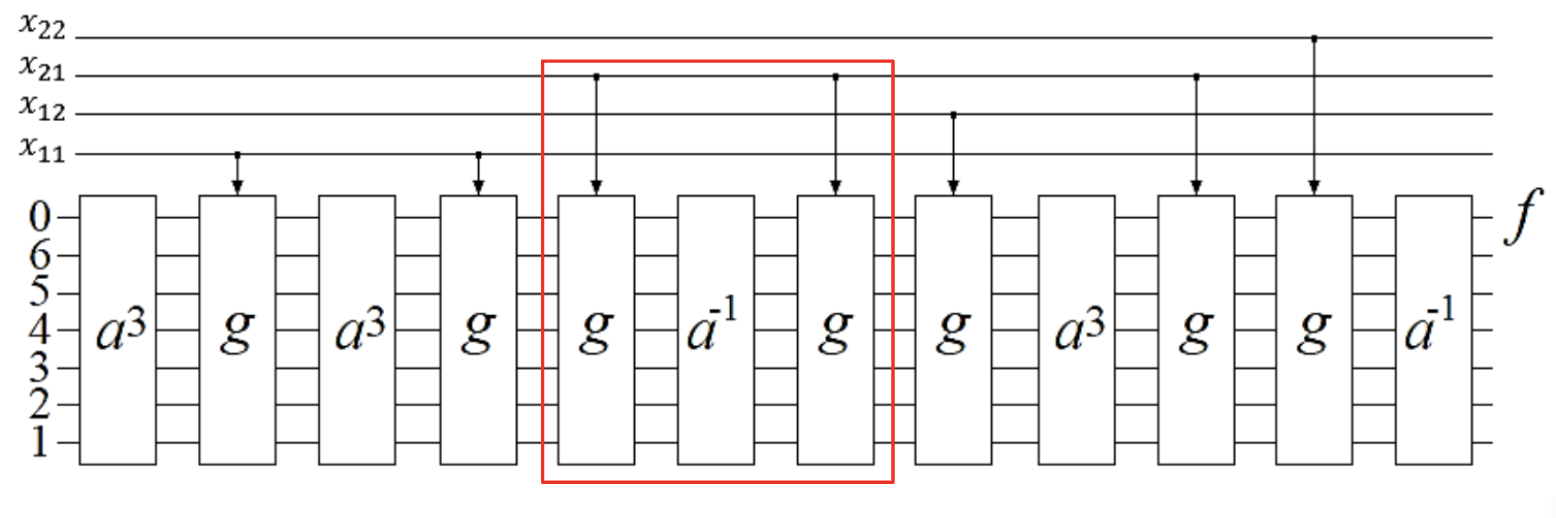}
    \caption{$g^{x_{21}}a^{-1}g^{x_{21}}$}
    \label{f49}
\end{figure}

When $x_{21} = 1$, $g^{x_{21}}a^{-1}g^{x_{21}} = a^1$. When $x_{21} = 0$, $g^{x_{21}}a^{-1}g^{x_{21}} = a^6$.  The truth table for $g^{x_{21}}a^{-1}g^{x_{21}}$ is shown in Figure \ref{f50}: 

\begin{figure}[!h]
    \centering
    \begin{tabular}{|c|c|c|c|c|c|c|c|}
    \textbf{$x_{21}$} & \textbf{$p$} & \textbf{$q$} & \textbf{$r$} & \textbf{$s$} & \textbf{$t$} & \textbf{$u$} & \textbf{$v$} \\

    \hline
    0 & 1 & 2 & 3 & 4 & 5 & 6 & 0 \\
    1 & 6 & 0 & 1 & 2 & 3 & 4 & 5
    \end{tabular}
    \caption{$g^{x_{21}}a^{-1}g^{x_{21}}$ Truth Table}
    \label{f50}
\end{figure}

The circuit for $g^{x_{21}}a^{-1}g^{x_{21}}$ is shown in Figure \ref{f51}.

\begin{figure}[!ht]
    \centering
    \scalebox{.7}{
    \begin{quantikz}
    & \ctrl{7} & \ctrl{7} & \ctrl{7} & \ctrl{7} & \ctrl{7} & \ctrl{7} & \targ{} & \ctrl{2} & \ctrl{3} & \ctrl{4} & \ctrl{5} & \ctrl{6} & \ctrl{7} & \qw \\
    &\swap{6} & \qw & \qw & \qw & \qw & \qw & \qw & \swap{1} & \qw & \qw  & \qw & \qw & \qw & \qw \\
    & \qw & \swap{5} & \qw & \qw & \qw & \qw & \qw & \swap{0} & \swap{1} & \qw & \qw & \qw & \qw & \qw \\
    & \qw & \qw & \swap{4} & \qw & \qw & \qw & \qw & \qw & \swap{0} & \swap{1} & \qw & \qw & \qw & \qw \\
    & \qw & \qw & \qw & \swap{3} & \qw & \qw & \qw & \qw & \qw & \swap{0} & \swap{1} & \qw & \qw & \qw \\
    & \qw & \qw & \qw & \qw & \swap{2} & \qw & \qw & \qw & \qw & \qw & \swap{0} & \swap{1} & \qw & \qw \\
    & \qw & \qw & \qw & \qw & \qw & \swap{1} & \qw & \qw & \qw & \qw & \qw & \swap{0} & \swap{1} & \qw \\
    & \swap{0} & \swap{0} & \swap{0} & \swap{0} & \swap{0} & \swap{0} & \qw & \qw & \qw & \qw & \qw & \qw & \swap{0} & \qw \\
    \end{quantikz}}
    \caption{$g^{x_{21}}a^{-1}g^{x_{21}}$ reduced circuit}
    \label{f51}
\end{figure}
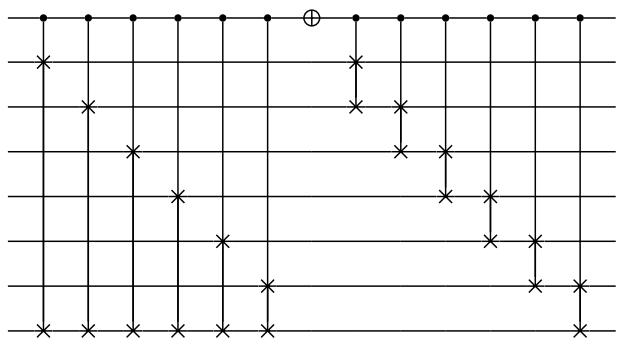

Next, notice that we can remove the final $a^{-1}$ gate by moving the output, $f(x)$, to the second wire.

\begin{figure}[!ht]
   \centering
    \includegraphics[width=8cm]{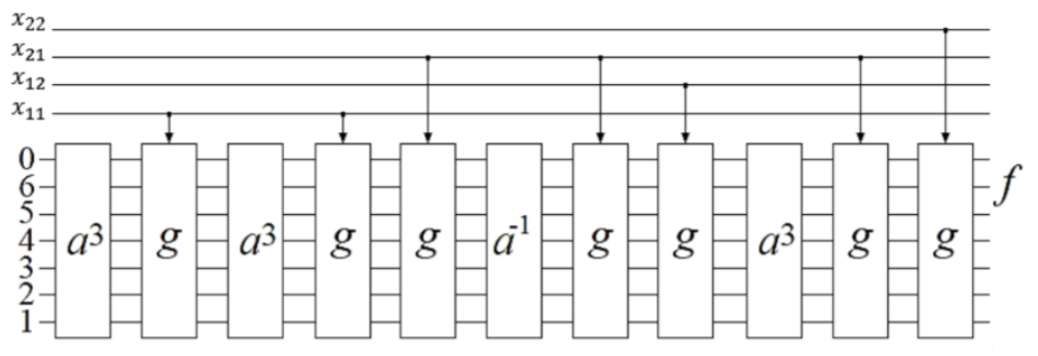}
    \caption{Reduced cascade after moving output to the second wire}
    \label{f52}
\end{figure}

Finally using our reduced circuits, we build the complete circuit for $f$ as shown in Figure \ref{f53}.

\begin{figure}[H]
    \centering
    \scalebox{.49}{
    \begin{quantikz}
    & \lstick{$x_{22}$} & \qw & \qw & \qw & \qw & \qw & \qw & \qw & \qw & \qw & \qw & \qw & \qw & \qw & \qw & \qw & \qw & \qw & \qw & \qw & \qw & \qw & \qw & \qw & \qw & \qw & \qw & \qw & \qw & \qw & \qw & \qw & \qw & \qw & \ctrl{10} & \ctrl{8} & \ctrl{9} & \qw \\
    & \lstick{$x_{21}$} & \qw & \qw & \qw & \qw & \qw & \qw & \qw & \ctrl{9} & \ctrl{9} & \ctrl{9} & \ctrl{9} & \ctrl{9} & \ctrl{9} & \targ{} & \ctrl{4} & \ctrl{5} & \ctrl{6} & \ctrl{7} & \ctrl{8} & \ctrl{9} & \targ{} & \qw & \qw & \qw & \qw & \qw & \qw & \qw & \qw & \qw & \ctrl{9} & \ctrl{8} & \ctrl{7} & \qw & \qw & \qw & \qw \\
    & \lstick{$x_{12}$} & \qw & \qw & \qw & \qw & \qw & \qw & \qw & \qw & \qw & \qw & \qw & \qw & \qw & \qw & \qw & \qw & \qw & \qw & \qw & \qw & \qw & \ctrl{8} & \ctrl{7} & \ctrl{6} & \qw & \qw & \qw & \qw & \qw & \qw & \qw & \qw & \qw & \qw & \qw & \qw & \qw \\
    & \lstick{$x_{11}$} & \targ{} & \ctrl{2} & \ctrl{3} & \ctrl{4} & \ctrl{5} & \ctrl{6} & \ctrl{7} & \qw & \qw & \qw & \qw & \qw & \qw & \qw & \qw & \qw & \qw & \qw & \qw & \qw & \qw & \qw & \qw & \qw & \qw & \qw & \qw & \qw & \qw & \qw & \qw & \qw & \qw & \qw & \qw & \qw \\
    & \lstick{0} & \qw & \swap{1} & \qw & \qw & \qw & \qw & \qw & \swap{6} & \qw & \qw & \qw & \qw & \qw & \qw & \swap{1} & \qw & \qw & \qw & \qw & \qw & \qw & \qw & \qw & \qw & \swap{4} & \qw & \qw & \qw & \qw & \qw & \qw & \qw & \qw & \qw & \qw & \qw & \qw  \\
    & \lstick{6} & \qw & \swap{0} & \swap{1} & \qw & \qw & \qw & \qw & \qw & \swap{5} & \qw & \qw & \qw & \qw & \qw & \swap{0} & \swap{1} & \qw & \qw & \qw & \qw & \qw & \swap{5} & \qw & \qw & \qw & \qw & \qw & \qw & \qw & \swap{3} & \swap{5} & \qw & \qw & \swap{5} & \qw & \qw & \qw \rstick{$f$} \\
    & \lstick{5} & \qw & \qw & \swap{0} & \swap{1} & \qw & \qw & \qw & \qw & \qw & \swap{4} & \qw & \qw & \qw & \qw & \qw & \swap{0} & \swap{1} & \qw & \qw & \qw & \qw & \qw & \swap{3} & \qw & \qw & \qw & \qw & \swap{2} & \qw & \qw & \qw & \swap{3} & \qw & \qw & \swap{3} & \qw & \qw \\
    & \lstick{4} & \qw & \qw & \qw & \swap{0} & \swap{1} & \qw & \qw & \qw & \qw & \qw & \swap{3} & \qw & \qw & \qw & \qw & \qw & \swap{0} & \swap{1} & \qw & \qw & \qw & \qw & \qw & \swap{1} & \qw & \swap{1} & \qw & \qw & \qw & \qw & \qw & \qw & \swap{1} & \qw & \qw & \swap{1} & \qw \\
    & \lstick{3} & \qw & \qw & \qw & \qw & \swap{0} & \swap{1} & \qw & \qw & \qw & \qw & \qw & \swap{2} & \qw & \qw & \qw & \qw & \qw & \swap{0} & \swap{1} & \qw & \qw & \qw & \qw & \swap{0} & \swap{0} & \swap{0} & \swap{2} & \swap{0} & \swap{1} & \swap{0} & \qw & \qw & \swap{0} & \qw & \qw & \swap{0} & \qw \\
    & \lstick{2} & \qw & \qw & \qw & \qw & \qw & \swap{0} & \swap{1} & \qw & \qw & \qw & \qw & \qw & \swap{1} & \qw & \qw & \qw & \qw & \qw & \swap{0} & \swap{1} & \qw & \qw & \swap{0} & \qw & \qw & \qw & \qw & \qw & \swap{0} & \qw & \qw & \swap{0} & \qw & \qw & \swap{0} & \qw & \qw \\
    & \lstick{2} & \qw & \qw & \qw & \qw & \qw & \qw & \swap{0} & \swap{0} & \swap{0} & \swap{0} & \swap{0} & \swap{0} & \swap{0} & \qw & \qw & \qw & \qw & \qw & \qw & \swap{0} & \qw & \swap{0} & \qw & \qw & \qw & \qw & \swap{0} & \qw & \qw & \qw & \swap{0} & \qw & \qw & \swap{0} & \qw & \qw & \qw \\
    \end{quantikz}}
    \caption{$f(x_{22}, x_{12}, x_{21}, x_{11}) = 2(x_{22} + x_{21}) + (x_{12} + x_{11})$  Circuit}
    \label{f53}
\end{figure}
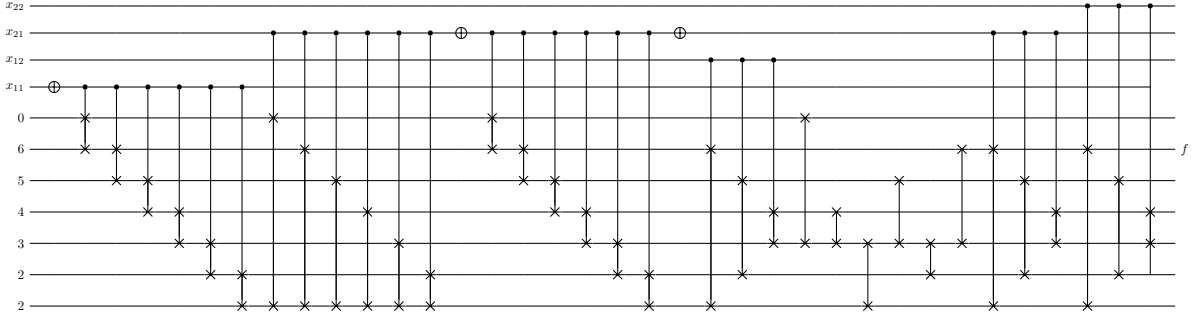

\textbf{Example 5:} Consider the binary input, modulo 5 output function $f$ with the following truth table:

\begin{figure}[!h]
    \centering
    \scalebox{0.9}{
    \begin{tabular}{|c|c|c|c|c|}
    \textbf{$x_{22}$} & \textbf{$x_{12}$} & \textbf{$x_{21}$} & \textbf{$x_{11}$} & \textbf{$\vec{F}$} \\

    \hline
    0 & 0 & 0 & 0 & 4\\
    0 & 0 & 0 & 1 & 3\\
    0 & 0 & 1 & 0 & 2\\
    0 & 0 & 1 & 1 & 0\\

    0 & 1 & 0 & 0 & 3\\
    0 & 1 & 0 & 1 & 4\\
    0 & 1 & 1 & 0 & 3\\
    0 & 1 & 1 & 1 & 1\\

    1 & 0 & 0 & 0 & 3\\
    1 & 0 & 0 & 1 & 0\\
    1 & 0 & 1 & 0 & 2\\
    1 & 0 & 1 & 1 & 4\\

    1 & 1 & 0 & 0 & 0\\
    1 & 1 & 0 & 1 & 4\\
    1 & 1 & 1 & 0 & 1\\
    1 & 1 & 1 & 1 & 4
    \end{tabular}}
    \caption{Function Truth Table}
    \label{f54}
\end{figure}

The truth vector is $\vec{F} = [4,3,2,0,3,4,3,1,3,0,2,4,0,4,1,4]^T$. Since ${W_4}^2 = 2^4I_4 \equiv I_4$ (mod 5),  ${W_4}^{-1} \equiv W_4$ (mod 5). Thus, \[\vec{w} = ({W_4}^{-1})\vec{F} \equiv W_4\vec{F} \text{ (mod 5)}\] \[= [3, 3, -1, 0, 3, 0, 0, 0, -1, 0, 0, 0, 0, 0, 0, 0]^T \text{ (mod 5)}\]

After simplification, the canonical cascade is the following: \[f = a^3g^{x_{11}}a^3g^{x_{11}}g^{x_{21}}a^{-1}g^{x_{21}}g^{x_{12}}a^3g^{x_{21}}g^{x_{22}}a^{-1}\]

The canonical cascade circuit diagram for $f$ is shown in Figure \ref{f55}. 

\begin{figure}[H]
    \centering
    \includegraphics[width=7.2cm]{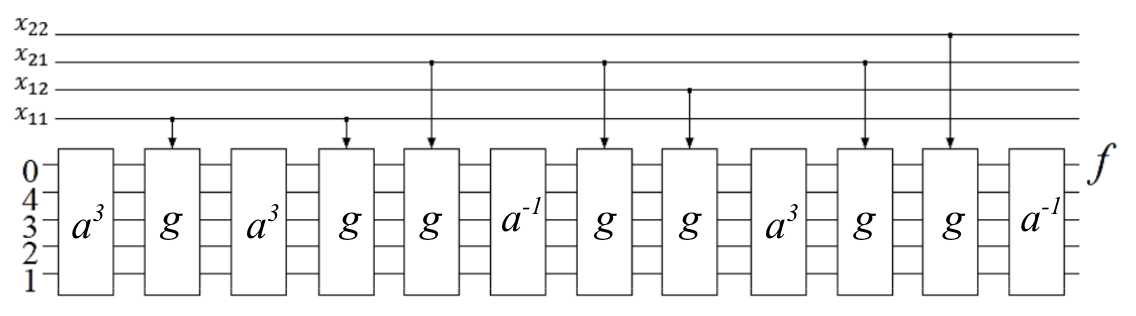}
    \caption{Reduced Canonical Cascade for $f$}
    \label{f55}
\end{figure}

\textbf{Local Transformations:} The next phase will be to apply local transformations to simplify the cascade. Consider the first four gates shown in Figure \ref{f56}. 

\begin{figure}[!ht]
    \centering
    \includegraphics[width=9.5cm]{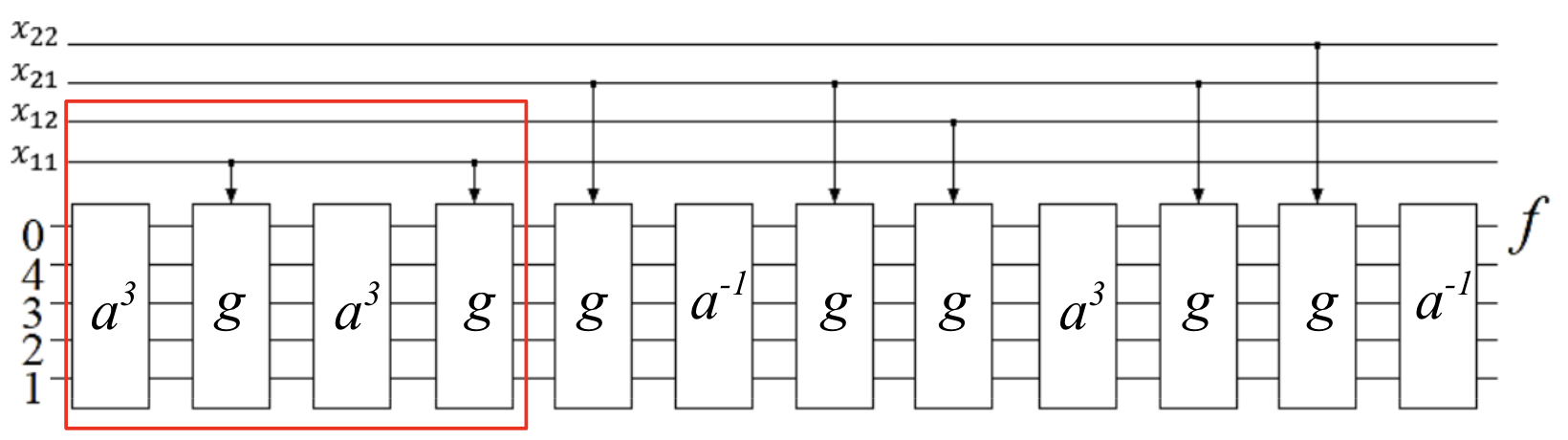}
    \caption{$g^{x_{11}}a^3g^{x_{11}}$}
    \label{f56}
\end{figure}

Since $gag = a^{-1}$ when the control qubit is 1, $a^3ga^3g = a^3(a^{-3}) = I$ when $x_{11} = 1$. When $x_{11}=0$, $a^3ga^3g = a^6 = a$. The circuit for $a^3g^{x_{11}}a^3g^{x_{11}}$ is shown in Figure \ref{f57}.

\begin{figure}[h]
    \centering
    \scalebox{0.9}{
    \begin{quantikz}
    & \lstick{$x_{11}$} & \targ{} & \ctrl{2} & \ctrl{4} & \ctrl{4} & \ctrl{5} & \qw \\
    & \lstick{0}  & \qw & \swap{4} & \qw & \qw  & \qw & \qw  \rstick{$p$} \\
    & \lstick{4}  & \qw & \qw & \swap{3} & \qw & \qw & \qw  \rstick{$q$} \\
    & \lstick{3}  & \qw & \qw & \qw & \swap{2} & \qw & \qw \rstick{$r$} \\
    & \lstick{2}  & \qw & \qw & \qw & \qw & \swap{1} & \qw \rstick{$s$} \\
    & \lstick{1}  & \qw & \swap{0} & \swap{0} & \swap{0} & \swap{0} & \qw \rstick{$t$} \\
    \end{quantikz}}
    \caption{$a^3g^{x_{11}}a^3g^{x_{11}}$ reduced circuit}
    \label{f57}
\end{figure}
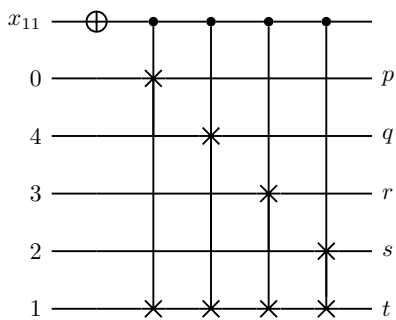

Similarly, we can simplify the three gates shown in Figure \ref{f58}. 

\begin{figure}[!ht]
    \centering
    \includegraphics[width=11cm]{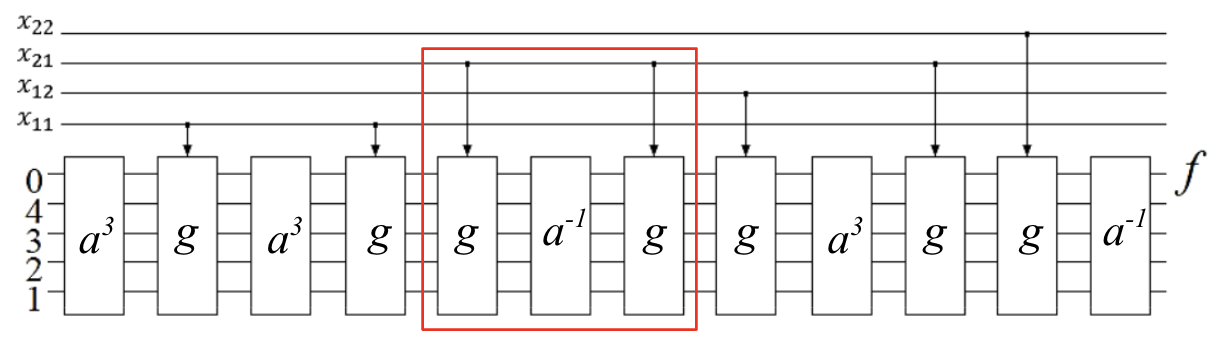}
    \caption{$g^{x_{21}}a^{-1}g^{x_{21}}$}
    \label{f58}
\end{figure}

The truth table for $g^{x_{21}}a^{-1}g^{x_{21}}$ is shown in Figure \ref{f59}: 

\begin{figure}[!h]
    \centering
    \begin{tabular}{|c|c|c|c|c|c|}
    \textbf{$x_{21}$} & \textbf{$p$} & \textbf{$q$} & \textbf{$r$} & \textbf{$s$} & \textbf{$t$} \\

    \hline
    0 & 4 & 3 & 2 & 1 & 0\\
    1 & 1 & 0 & 4 & 3 & 2
    \end{tabular}
    \caption{$g^{x_{21}}a^{-1}g^{x_{21}}$ Truth Table}
    \label{f59}
\end{figure}

The circuit for $g^{x_{21}}a^{-1}g^{x_{21}}$ is shown in Figure \ref{f60}.

\begin{figure}[H]
    \centering
    \scalebox{0.65}{
    \begin{quantikz}
    & \lstick{$x_{21}$} & \ctrl{4} & \ctrl{4} & \ctrl{5} & \ctrl{5} & \targ{} & \ctrl{2} & \ctrl{3} & \ctrl{4} & \ctrl{5} & \qw \\
    & \lstick{0}  & \swap{4} & \qw & \qw  & \qw & \qw & \swap{1} & \qw & \qw & \qw & \qw \rstick{$p$} \\
    & \lstick{4}  & \qw & \swap{3} & \qw & \qw & \qw & \swap{0} & \swap{1} & \qw & \qw & \qw \rstick{$q$} \\
    & \lstick{3}  & \qw & \qw & \swap{2} & \qw & \qw & \qw & \swap{0} & \swap{1} & \qw & \qw \rstick{$r$} \\
    & \lstick{2}  & \qw & \qw & \qw & \swap{1} & \qw & \qw & \qw & \swap{0} & \swap{1} & \qw \rstick{$s$} \\
    & \lstick{1}  & \swap{0} & \swap{0} & \swap{0} & \swap{0} & \qw & \qw & \qw & \qw & \swap{0} & \qw \rstick{$t$} \\
    \end{quantikz}}
    \caption{$g^{x_{21}}a^{-1}g^{x_{21}}$ reduced circuit}
    \label{f60}
\end{figure}

We can also remove the last $a^{-1}$ gate and move the output to the second wire. Using our reduced circuits, we build the final circuit for $f$ as shown in Figure \ref{f61}.

\begin{figure}[h]
    \centering
    \scalebox{.65}{
    \begin{quantikz}
    & \lstick{$x_{22}$}  & \qw & \qw & \qw & \qw & \qw & \qw & \qw & \qw & \qw & \qw & \qw & \qw & \qw & \qw & \qw & \qw & \qw & \qw & \qw & \qw & \qw & \qw & \qw & \qw & \qw & \qw & \qw & \qw & \ctrl{6} & \ctrl{7} & \qw  \\
     & \lstick{$x_{21}$} & \qw & \qw & \qw & \qw & \qw & \qw & \qw & \qw & \qw  & \qw & \ctrl{6} & \ctrl{6} & \ctrl{7} & \ctrl{7} & \targ{} & \ctrl{4} & \ctrl{5} & \ctrl{6} & \ctrl{7} & \qw & \qw & \qw & \qw & \qw & \qw & \targ{} & \ctrl{5} & \ctrl{6} & \qw & \qw & \qw  \\
    & \lstick{$x_{12}$} & \qw & \qw & \qw & \qw & \qw & \qw & \qw & \qw & \qw & \qw & \qw & \qw & \qw & \qw & \qw & \qw & \qw & \qw & \qw & \ctrl{4} & \ctrl{5} & \qw & \qw & \qw & \qw & \qw & \qw & \qw & \qw & \qw & \qw \\
    & \lstick{$x_{11}$}  & \ctrl{3} & \ctrl{4} & \ctrl{5} & \ctrl{3} & \targ{} & \ctrl{2} & \ctrl{4} & \ctrl{4} & \ctrl{4} & \qw & \qw & \qw & \qw & \qw & \qw & \qw & \qw & \qw & \qw & \qw & \qw & \qw & \qw & \qw & \qw & \qw & \qw & \qw & \qw & \qw & \qw \\
    & \lstick{2}  & \swap{3} & \qw & \qw  & \qw & \qw & \swap{2} & \qw & \qw & \qw & \qw & \swap{4} & \qw & \qw  & \qw & \qw & \swap{1} & \qw & \qw & \qw & \qw & \qw & \swap{2} & \qw & \qw & \qw & \qw & \qw & \qw & \qw & \qw & \qw \rstick{$p$} \\
    & \lstick{1}  & \qw & \qw & \qw & \swap{2} & \qw & \qw & \qw & \qw & \swap{3} & \qw & \qw & \swap{3} & \qw & \qw & \qw & \swap{0} & \swap{1} & \qw & \qw & \swap{3} & \qw & \qw & \qw & \qw & \swap{3} & \qw & \swap{3} & \qw & \swap{3} & \qw & \qw \rstick{$q = \mathit{f}$} \\
    & \lstick{0}  & \qw & \swap{1} & \qw & \qw & \qw & \swap{0} & \swap{1} & \swap{2} & \qw & \qw & \qw & \qw & \swap{2} & \qw & \qw & \qw & \swap{0} & \swap{1} & \qw & \qw & \swap{1} & \swap{0} & \swap{1} & \swap{2} & \qw & \qw & \qw & \swap{1} & \qw & \swap{1} & \qw \rstick{$r$} \\
    & \lstick{4}  & \swap{0} & \swap{0} & \swap{1} & \swap{0} & \qw & \qw & \swap{0} & \qw & \qw & \qw & \qw & \qw & \qw & \swap{1} & \qw & \qw & \qw & \swap{0} & \swap{1} & \qw & \swap{0} & \qw & \swap{0} & \qw & \qw & \qw & \qw & \swap{0} & \qw & \swap{0} & \qw \rstick{$s$} \\
    & \lstick{3}  & \qw & \qw & \swap{0} & \qw & \qw & \qw & \qw & \swap{0} & \swap{0} & \qw & \swap{0} & \swap{0} & \swap{0} & \swap{0} & \qw & \qw & \qw & \qw & \swap{0} & \swap{0} & \qw & \qw & \qw & \swap{0} & \swap{0} & \qw & \swap{0} & \qw & \swap{0} & \qw & \qw \rstick{$t$} \\
    \end{quantikz}}
    \caption{Circuit for $f$}
    \label{f61}
\end{figure}

\textbf{Example 6: } Consider the function $f(x_{1}, x_{2}, x_{3}, x_{4}) = 2(x_{1} + x_{3}) + (x_{2} + x_{4})$, the two-bit, four binary input variable, modulo 7 output adder from example 5. We propose another method to realize this function using the group $D_3$ with multiple outputs. The map of this function is shown in Figure \ref{f62}:

\begin{figure}[h!t]
\begin{center}
    \begin{tikzpicture}[scale = 0.8, karnaugh, American style, kmlabel left/.style={black, left=0pt}, kmlabel top/.style={black, yshift=8pt}]
  \karnaughmaptabvert{4}{}{{$x_3$}{$x_4$}{$x_1$}{$x_2$}}%
     {{0}{1}
      {3}{2}
      {1}{2}
      {4}{3}
      {3}{4}
      {6}{5}
      {2}{3}
      {5}{4}
      }{}
\end{tikzpicture}
\caption{Map for function $f(x_{1}, x_{2}, x_{3}, x_{4}) = 2(x_{1} + x_{3}) + (x_{2} + x_{4})$}
\label{f62}
\end{center}
\end{figure}

The values from $0$ to $6$ can be written as follows in ternary:

\[0 - 00_3\]
\[1  - 01_3\]
\[2 - 02_3\]
\[3 - 10_3\]
\[4 - 11_3\]
\[5 - 12_3\]
\[6 - 20_3\]

Replacing the values in our map in Figure \ref{f62}, we obtain Figure \ref{f63}:

\begin{figure}[H]
\begin{center}
    \begin{tikzpicture}[karnaugh, American style, kmlabel left/.style={black, left=0pt}, kmlabel top/.style={black, yshift=8pt}]
  \karnaughmaptabvert{4}{}{{$x_3$}{$x_4$}{$x_1$}{$x_2$}}%
     {{00}{01}
      {10}{02}
      {01}{02}
      {11}{10}
      {10}{11}
      {20}{12}
      {02}{10}
      {12}{11}
      }{}
\end{tikzpicture}
\caption{Ternary map for function $f(x_{1}, x_{2}, x_{3}, x_{4}) = 2(x_{1} + x_{3}) + (x_{2} + x_{4})$}
\label{f63}
\end{center}
\end{figure}

We can split this map into two by considering the first and second digits separately:

\begin{figure}[!ht]
\centering
\begin{subfigure}{.5\textwidth}
  \centering
   \begin{tikzpicture}[karnaugh, American style, kmlabel left/.style={black, left=0pt}, kmlabel top/.style={black, yshift=8pt}]
  \karnaughmaptabvert{4}{}{{$x_3$}{$x_4$}{$x_1$}{$x_2$}}%
     {{0}{0}
      {1}{0}
      {0}{0}
      {1}{1}
      {1}{1}
      {2}{1}
      {0}{1}
      {1}{1}
      }{}
\end{tikzpicture}
  \caption{Submap of First Digit}
\end{subfigure}%
\begin{subfigure}{.5\textwidth}
  \centering
   \begin{tikzpicture}[karnaugh, American style, kmlabel left/.style={black, left=0pt}, kmlabel top/.style={black, yshift=8pt}]
  \karnaughmaptabvert{4}{}{{$x_3$}{$x_4$}{$x_1$}{$x_2$}}%
     {{0}{1}
      {0}{2}
      {1}{2}
      {1}{0}
      {0}{1}
      {0}{2}
      {2}{0}
      {2}{1}
      }{}
\end{tikzpicture}
  \caption{Submap of Second Digit}
\end{subfigure}
\caption{Two submaps of ternary map of $f$}
\label{f64}
\end{figure}

We will now compute the canonical expression for submap $a$ from Figure \ref{f64}. The truth vector of submap a is [0  0  0  1  0  0  1  1  0  1  1  1  1  1  1  2]$^T$. After simplification, the canonical expression is the following: \[ g^{x_3+x_4}a^1g^{x_2+x_4}a^1g^{x_4}a^1g^{x_1+x_2+x_3}a^1g^{x_3+x_4}a^1g^{x_4}a^1g^{x_2+x_3+x_4}a^1g^{x_4}a^1g^{x_3+x_4}a^1g^{x_4}a^2  \]
 
Similarly, the truth vector for submap $b$ is [0  1  2  0  1  2  0  1  2  0  1  2  0  1  2  0]$^T$. After simplification, the canonical expression is the following: 
\[ g^{x_4}a^1g^{x_3+x_4}a^2g^{x_2+x_3}a^1g^{x_4}a^2g^{x_3}a^1g^{x_1+x_2+x_3+x_4}a^2g^{x_3}a^1g^{x_4}a^2g^{x_2+x_3}a^1g^{x_3+x_4}a^2  \]

Using these two expressions, we can create the cascade and circuit for the function.

}

\section{Upper Bound on Number of Gates}

{\parindent0pt

\noindent \emph{Lemma 1: }The maximum number of cells in an arbitrary $n$-variable input cascade  is $3*2^n - 4 - n$. 

\begin{proof}
Recall that \[F(x_1,x_2, \ldots x_n, x_{n+1}) = F_a(x_2, \ldots x_{n+1})g^{x_1}F_b(x_2, \ldots x_{n+1})g^{x_1}\]

Let $A(n)$ denote the number of $a$ gates in the n-variable input cascade before reduction. Then,

\[A(n+1) = 2A(n)\]
We can solve this recurrence relation using its characteristic equation [\ref{58}]:
\[\lambda-2 = 0\]
\[\Rightarrow A(n) = k2^n\]
Since $A(1) = 2,$ we have $k = 1$, so $A(n) = 2^k$

\medskip

Let $G(n)$ denote the number of $g$ gates in the n-variable input cascade before reduction. Then,

\[G(n+1) = 2G(n) + 2\]
This recurrence relation has characteristic equation:
\[\lambda^2 - 3\lambda + 2 = 0\]
\[(\lambda -1)(\lambda-2) = 0\]
\[\Rightarrow G(n) = k_12^n + k_21^n\]

Since $G(1) = 2, G(2) = 6,$ we have $k_1 = 2, k_2 = -2$, so $G(n) = 2*2^n - 2$.

\medskip

Note that the first and last $a$ gates and the last $g$ gates can always be removed to reduce the cascade. Thus, after reduction, the maximum number of $a$ gates is $2^n - 2$ and the maximum number of $g$ gates is $2*2^n - 2 - n$. Hence, the total number of gates in the cascade is $3*2^n - 4 - n$.

\end{proof}

\bigskip

\noindent Note that for a $k$-valued output function, each $a$ gate can be realized using $k-1$ SWAP gates and each $g$ gate can be realized using $\frac{k-1}{2}$ multivalued Fredkin Gates. Hence, by Lemma 1 the maximum number of SWAP or Fredkin gates in an $n$-variable input with $k$-valued output quantum circuit is $(k-1)(2^n-2) + \frac{k-1}{2}(2*2^n - 2 - n)$. More specifically, with a maximum of $(k-1)(2^n-2)$ SWAP gates and $\frac{k-1}{2}(2*2^n - 2 - n)$ Fredkin gates.

}

\section{Conclusion}

In the past, several authors developed methods to use group theory to design classical binary logical cascades [2, 9, 10, 16]. In this paper, we have extended the decomposition from [2, 3] and have created binary input and multivalued output quantum cascades using NOT, SWAP, and Controlled SWAP gates. The choice of our gates was determined by what is currently implementable on optical technologies, and the design of our cascades was motivated by practical quantum layouts. We have also created a method to realize a function with different valued outputs and have developed seven local transformations to simplify the final cascade circuits. Through these simplifications, we have shown that an arbitrary $n$-variable input cascade has a maximum of $3*2^n - 4-n$ cells, and the maximum number of individual multivalued Fredkin and SWAP gates in an arbitrary $n$-variable input and $k$-valued output function is $(k-1)(2^n-2) + \frac{k-1}{2}(2*2^n - 2 - n)$.

\section{References}

\begin{enumerate}
    
    \item Litwin, Przemysław, et al. "Ternary logic in the optical controlled-SWAP gate based on Laguerre-Gaussian modes of light." \textit{Optics Express} 32.9 (2024): 15258-15268. \label{32}
    
    \item Hurst. "Multiple-valued logic—Its status and its future." \textit{IEEE Transactions on Computers} 100.12 (1984): 1160-1179. \label{33}
    
    \item Smith, Kenneth C. "A multiple valued logic: a tutorial and appreciation." \textit{Computer} 21.4 (1988): 17-27. \label{34}

    \item Bérut, Antoine, et al. "Experimental verification of Landauer’s principle linking information and thermodynamics." \textit{Nature} 483.7388 (2012): 187-189. \label{50}

    \item Landauer, Rolf. "Dissipation and noise immunity in computation and communication." \textit{Nature} 335.6193 (1988): 779-784. \label{36}

    \item Landauer, Rolf. "Irreversibility and heat generation in the computing process." \textit{IBM Journal of Research and Development} 5.3 (1961): 183-191. \label{51}

    \item Hamerly, Ryan, et al. "Large-scale optical neural networks based on photoelectric multiplication." \textit{Physical Review X} 9.2 (2019): 021032. \label{47}

    \item Li, Jingming, et al. "Energy consumption of cryptocurrency mining: A study of electricity consumption in mining cryptocurrencies." \textit{Energy} 168 (2019): 160-168. \label{64}

    \item Pelucchi, Emanuele, et al. "The potential and global outlook of integrated photonics for quantum technologies." \textit{Nature Reviews Physics} 4.3 (2022): 194-208. \label{59}

    \item Chattopadhyay, Tanay. "All-optical modified Fredkin gate." \textit{IEEE Journal of Selected Topics in Quantum Electronics} 18.2 (2011): 585-592. \label{7}

    \item Ringbauer, Martin, et al. "A universal qudit quantum processor with trapped ions." \textit{Nature Physics} 18.9 (2022): 1053-1057. \label{30}

    \item Imai, Yoh, and Yoshihiro Ohtsuka. "Optical multiple-output and multiple-valued logic operation based on fringe shifting techniques using a spatial light modulator." \textit{Applied Optics} 26.2 (1987): 274-277. \label{48}
    
    \item Yi, Jin, He Huacan, and Lü Yangtian. "Ternary optical computer architecture." \textit{Physica Scripta} 2005.T118 (2005): 98. \label{49}
            
    \item Chattopadhyay, Tanay. "All-optical symmetric ternary logic gate." \textit{Optics \& Laser Technology} 42.6 (2010): 1014-1021. \label{54}
    
    \item Mukhopadhyay, Sourangshu. "Role of optics in super-fast information processing." \textit{Indian Journal of Physics} 84 (2010): 1069-1074. \label{55}
    
    \item Hung, William NN, et al. "Optimal synthesis of multiple output boolean functions using a set of quantum gates by symbolic reachability analysis." \textit{IEEE Transactions on Computer-Aided Design of Integrated Circuits and Systems} 25.9 (2006): 1652-1663. \label{1}

    \item Saraivanov, Miroslav, and Marek Perkowski. "Multi-valued Quantum Cascade Realization with Group Decomposition." \textit{2018 IEEE 48th International Symposium on Multiple-Valued Logic (ISMVL)}. IEEE, 2018. \label{3}
    
    \item Saraivanov, Michael S. Quantum Circuit Synthesis using Group Decomposition and Hilbert Spaces. MS thesis. Portland State University, 2013. \label{4}
    
    \item Miller, D. Michael, and Gerhard W. Dueck. "Search-based transformation synthesis for 3-valued reversible circuits." \textit{Reversible Computation: 12th International Conference, RC 2020, Oslo, Norway, July 9-10, 2020, Proceedings 12}. Springer International Publishing, 2020. \label{5}
    
    \item Soeken, Mathias, et al. "An extension of transformation-based reversible and quantum circuit synthesis." \textit{2016 IEEE International Symposium on Circuits and Systems (ISCAS)}. IEEE, 2016. \label{6}
    
    \item Hawash, Maher, and Marek Perkowski. "Using Hasse diagrams to synthesize ternary quantum circuits." \textit{2012 IEEE 42nd International Symposium on Multiple-Valued Logic}. IEEE, 2012. \label{17}
    
    \item Moraga, Claudio. "Aspects of Reversible and Quantum Computing in a $ p $-Valued Domain." \textit{IEEE Journal on Emerging and Selected Topics in Circuits and Systems} 6.1 (2016): 44-52. \label{18}
    
    \item Khan, Mozammel HA, Himanshu Thapliyal, and Edgard Munoz-Coreas. "Automatic synthesis of quaternary quantum circuits." \textit{The Journal of Supercomputing} 73 (2017): 1733-1759. \label{19}

    \item Di, Yao-Min, and Hai-Rui Wei. "Synthesis of multivalued quantum logic circuits by elementary gates." \textit{Physical Review A—Atomic, Molecular, and Optical Physics} 87.1 (2013): 012325. \label{44}
    
    \item Sasao, Tsutomu. "Cascade realizations of two-valued input multiple-valued output functions using decomposition of group functions." \textit{33rd International Symposium on Multiple-Valued Logic, 2003. Proceedings.} IEEE, 2003. \label{2}

    \item Elspas, Bernard, and Harold S. Stone. "Decomposition of group functions and the synthesis of multirail cascades." \textit{8th Annual Symposium on Switching and Automata Theory (SWAT 1967)}. IEEE, 1967. \label{9}

    \item Yoeli, Michael, and James Turner. "Decompositions of group functions with applications to two-rail cascades." \textit{Information and Control} 10.6 (1967): 565-571. \label{10}

    \item Delsarte, Philippe, and Jean-Jacques Quisquater. "Permutation cascades with normalized cells." \textit{Information and Control} 23.4 (1973): 344-356. \label{16}

    \item Bhattacharya, Animesh, Goutam K. Maity, and Amal K. Ghosh. "Optical quadruple Toffoli and Fredkin gate using SLM and Savart plate." \textit{Computational Intelligence, Communications, and Business Analytics: First International Conference, CICBA 2017, Kolkata, India, March 24–25, 2017, Revised Selected Papers, Part I}. Springer Singapore, 2017. \label{29}

    \item Milburn, Gerard J. "Quantum optical Fredkin gate." \textit{Physical Review Letters} 62.18 (1989): 2124. \label{37}

    \item O'Brien, Jeremy L., et al. "Demonstration of an all-optical quantum controlled-NOT gate." \textit{Nature} 426.6964 (2003): 264-267. \label{45}
    
    \item Lopes, J. H., et al. "Linear optical CNOT gate with orbital angular momentum and polarization." \textit{Quantum Information Processing} 18 (2019): 1-10. \label{46}

    \item Patel, Raj B., et al. "A quantum Fredkin gate." \textit{Science Advances} 2.3 (2016): e1501531. \label{52}
    
    \item Wang, Feiran, et al. "Experimental demonstration of a quantum controlled-SWAP gate with multiple degrees of freedom of a single photon." \textit{Quantum Science and Technology} 6.3 (2021): 035005. \label{53}
    
    \item Garai, Sisir Kumar. "A novel method of developing all optical frequency encoded Fredkin gates." \textit{Optics Communications} 313 (2014): 441-447. \label{56}

    \item Perkowski, Marek, et al. "Synthesis of quantum circuits in linear nearest neighbor model using positive Davio lattices." \textit{Facta Universitatis-Series: Electronics and Energetics} (2011). \label{20}

    \item Whitney, Mark, et al. "Automated generation of layout and control for quantum circuits." Proceedings of the 4th International Conference on Computing Frontiers. 2007. \label{15}

    \item Lukac, Martin, et al. "Geometric Refactoring of Quantum and Reversible Circuits: Quantum Layout." \textit{2020 23rd Euromicro Conference on Digital System Design (DSD)}. IEEE, 2020. \label{23}

    \item Tan, Bochen, and Jason Cong. "Optimal layout synthesis for quantum computing." \textit{Proceedings of the 39th International Conference on Computer-Aided Design}. 2020. \label{24}

    \item Di, Yao-Min, and Hai-Rui Wei. "Elementary gates for ternary quantum logic circuit." arXiv preprint arXiv:1105.5485 (2011). \label{11}
    
    \item Wang, Yuchen, et al. "Qudits and high-dimensional quantum computing." \textit{Frontiers in Physics} 8 (2020): 589504. \label{12}
        
    \item Mandal, Dhoumendra, Sumana Mandal, and Sisir Kumar Garai. "Alternative approach of developing all-optical Fredkin and Toffoli gates." \textit{Optics \& Laser Technology} 72 (2015): 33-41. \label{26}
    
    \item Feinstein, David Y., and Mitchell A. Thornton. "Using the Asynchronous Paradigm for Reversible Sequential Circuit Implementation." 2\textit{012 IEEE 42nd International Symposium on Multiple-Valued Logic}. IEEE, 2012. \label{27}
    
    \item Picton, P. D. "Fredkin gates as a basis for comparison of different logic design solutions." \textit{IEE Colloquium on Synthesis and Optimisation of Logic Systems}. IET, 1994. \label{28}

    \item Wang, B., and L-M. Duan. "Implementation scheme of controlled SWAP gates for quantum fingerprinting and photonic quantum computation." \textit{Physical Review A—Atomic, Molecular, and Optical Physics} 75.5 (2007): 050304. \label{35}

    \item Shamir, Joseph, et al. "Optical computing and the Fredkin gates." \textit{Applied Optics} 25.10 (1986): 1604-1607. \label{38}
    
    \item Cuykendall, Robert, and Debra McMillin. "Control-specific optical Fredkin circuits." \textit{Applied Optics} 26.10 (1987): 1959-1963. \label{39}
    
    \item Mirsalehi, Mir M., Joseph Shamir, and H. John Caulfield. "Residue arithmetic processing utilizing optical Fredkin gate arrays." \textit{Applied Optics} 26.18 (1987): 3940-3946. \label{40}
    
    \item Poustie, A. J., and K. J. Blow. "Demonstration of an all-optical Fredkin gate." \textit{Optics Communications} 174.1-4 (2000): 317-320. \label{41}

    \item S. Kotiyal, Saurabh, Himanshu Thapliyal, and Nagarajan Ranganathan. "Design of a ternary barrel shifter using multiple-valued reversible logic." \textit{10th IEEE International Conference on Nanotechnology}. IEEE, 2010. \label{42}
    
    \item Deibuk, Vitaly, Iryna Turchenko, and Vladyslav Shults. "Optimized design of the universal ternary gates for quantum/reversible computing." \textit{2015 IEEE 8th International Conference on Intelligent Data Acquisition and Advanced Computing Systems: Technology and Applications (IDAACS)}. Vol. 2. IEEE, 2015. \label{43}

    \item Khan, Faisal Shah, and Marek Perkowski. "Synthesis of multi-qudit hybrid and d-valued quantum logic circuits by decomposition." \textit{Theoretical Computer Science} 367.3 (2006): 336-346. \label{60}

    \item Muthukrishnan, Ashok, and Carlos R. Stroud Jr. "Multivalued logic gates for quantum computation." \textit{Physical review A} 62.5 (2000): 052309. \label{61}

    \item O'brien, Jeremy L. "Optical quantum computing." \textit{Science} 318.5856 (2007): 1567-1570. \label{62}

    \item Ralph, T. C., K. J. Resch, and Alexei Gilchrist. "Efficient Toffoli gates using qudits." \textit{Physical Review A—Atomic, Molecular, and Optical Physics} 75.2 (2007): 022313. \label{63}

    \item Pedersen, John. “Groups of Small Order.” Groups of Small Order, mathweb.ucsd.edu/~atparris/small\_groups.html. Accessed 3 Dec. 2024.  \label{14}

    \item Ghosh, Swaroop, Swarup Bhunia, and Kaushik Roy. "Shannon expansion based supply-gated logic for improved power and testability." \textit{14th Asian Test Symposium (ATS'05)}. IEEE, 2005. \label{31}
    
    \item Falkowski, Bogdan J., and Marek A. Perkowski. "Walsh Type Transforms for Completely and Incompletely Specified Multiple-Valued Input Binary Functions." \textit{ISMVL}. 1990.\label{8}
    
    \item Falkowski, Bogdan J., Ingo Schafer, and Marek A. Perkowski. "Effective computer methods for the calculation of Rademacher-Walsh spectrum for completely and incompletely specified Boolean functions." \textit{IEEE Transactions on Computer-Aided Design of Integrated Circuits and Systems} 11.10 (1992): 1207-1226. \label{13}
    
    \item Falkowski, Bogdan J. "Properties and ways of calculation of multi-polarity generalized Walsh transforms." \textit{IEEE Transactions on Circuits and Systems II: Analog and Digital Signal Processing} 41.6 (1994): 380-391. \label{21}
    
    \item Thornton, Mitchell Aaron, Rolf Drechsler, and D. Michael Miller. \textit{Spectral Techniques in VLSI CAD}. Springer Science \& Business Media, 2012. \label{22}
        
   \item Henderson, Keith W. "Some notes on the Walsh functions." \textit{IEEE Transactions on Electronic Computers} 1 (1964): 50-52. \label{57}
    
    \item Martin, George Edward. \textit{Counting: The Art of Enumerative Combinatorics}. New York: Springer, 2001.\label{58}

\end{enumerate}

\end{document}